\documentclass[prd,10pt,twocolumn,showpacs,nofootinbib,aps,superscriptaddress,preprintnumbers]{revtex4-1}

 \usepackage{bm}
 \usepackage{graphicx}
 \usepackage{multirow}
 \usepackage{amsmath}
 \usepackage{mathrsfs}
 \usepackage{amssymb}
 \usepackage{hyperref}
 
\usepackage{upgreek}
 \usepackage{yfonts}
 \usepackage{color}
 \usepackage{xspace}
 \usepackage{verbatim}
 \usepackage{mathtools}
 \usepackage{booktabs}
 \usepackage{tabularx}
\usepackage[normalem]{ulem}
 \usepackage[utf8]{inputenc}
 \usepackage[dvipsnames,table]{xcolor}
 \hypersetup{urlcolor=BlueViolet,
             citecolor=WildStrawberry,
             linkcolor=PineGreen, 
             colorlinks=true}
 \definecolor{lightgray}{rgb}{0.9,0.9,0.9}	    
 \definecolor{green}{rgb}{0,0.5,0}
 \definecolor{red}{rgb}{1,0,0}
 \definecolor{blue}{rgb}{0,0,0.5}

 \newcommand{\eV}{\text{e\kern-0.15ex V}\xspace}
 \newcommand{\keV}{\text{k\eV}\xspace}
 
 \newcommand{\GeV}{\text{G\eV}\xspace}
 \newcommand{\TeV}{\text{T\kern-0.1ex \eV}\xspace}
 \newcommand{\epsa}{\hat{\boldsymbol{\epsilon}}_1}
 \newcommand{\epsb}{\hat{\boldsymbol{\epsilon}}_2}

 \newcommand{\vlab}{\mathbf{v}_\mathrm{lab}}
 
 \newcommand{\vmin}{v_{\rm min}}
 
 \newcommand{\vesc}{v_\mathrm{esc}}
 \newcommand{\kms}{\textrm{ km \!s}^{-1}}

 \newcommand{\dbd}[2]{\ifmmode \frac{\textrm{d}#1}{\textrm{d}#2}\else $\textrm{d}#1/\textrm{d}#2$\fi}
 \newcommand{\pbp}[2]{\ifmmode \frac{\partial#1}{\partial#2}\else $\partial#1/\partial#2$\fi}
 \newcommand{\erf}{\mathrm{erf}}
 \newcommand{\erfi}{\mathrm{erfi}}
 \newcommand{\ra}[1]{\renewcommand{\arraystretch}{#1}}

 \newcommand{\vbf}{\mathbf{v}}

 \DeclareMathAlphabet{\mathpzc}{OT1}{pzc}{m}{it}

 \newcommand{\fR}{f_{\rm R}}
 \newcommand{\fS}{f_{\rm S}}
 
 \newcommand{\SHMpp}{SHM$^{++}$\xspace}
 \newcommand{\Gaia}{\!{\it Gaia}\xspace}

 \newcommand{\be}{\begin{equation}}
 \newcommand{\ee}{\end{equation}}
 \newcommand{\bea}{\begin{eqnarray}}
 \newcommand{\eea}{\end{eqnarray}}

\begin{document}

 \title{Dark Shards: velocity substructure from \Gaia and direct searches for dark matter}

 \author{Ciaran A. J. O'Hare} \email{ciaran.aj.ohare@gmail.com}
 \affiliation{The University of Sydney, School of Physics, New South Wales, 2006, Australia}

 \author{N. Wyn Evans} \email{nwe@ast.cam.ac.uk}
 \affiliation{Institute of Astronomy, Madingley Rd, Cambridge, CB3 0HA, United Kingdom}

 \author{Christopher McCabe} \email{christopher.mccabe@kcl.ac.uk}
 \affiliation{Department of Physics, King's College London, Strand, London, WC2R 2LS, United Kingdom}

 \author{GyuChul Myeong} \email{gm564@cam.ac.uk}
 \affiliation{Institute of Astronomy, Madingley Rd, Cambridge, CB3 0HA, United Kingdom}

 \author{Vasily Belokurov}
 \affiliation{Institute of Astronomy, Madingley Rd, Cambridge, CB3 0HA, United Kingdom}

 \preprint{KCL-PH-TH-2019-58}

 \date{\today}
 \smallskip
 \begin{abstract}
Data from the \Gaia satellite show that the solar neighbourhood of the
Milky Way's stellar halo is imprinted with substructure from several
accretion events.  Evidence of these events is found in ``the
Shards'', stars clustering with high significance in both action space
and metallicity.  Stars in the Shards share a common origin, likely as
ancient satellite galaxies of the Milky Way, so will be embedded in
dark matter (DM) counterparts.  These ``Dark Shards'' contain two
substantial streams (S1 and S2), as well as several retrograde,
prograde and lower energy substructures.  The retrograde stream S1 has a
very high Earth-frame speed of $\sim 550 \kms$ while S2 moves on a
prograde, but highly polar orbit and enhances the peak of the speed
distribution at around $300 \kms$.  The presence of the Dark Shards
locally leads to modifications of many to the fundamental properties
of experimental DM signals.  The~S2 stream in particular gives rise to
an array of effects in searches for axions and in the time dependence
of nuclear recoils: shifting the peak day, inducing nonsinusoidal
distortions, and increasing the importance of the gravitational
focusing of DM by the Sun.  Dark Shards additionally bring new
features for directional signals, while also enhancing the DM flux
towards Cygnus.
 \end{abstract}

 \maketitle

 \section{Introduction}

 The stellar populations of the Milky Way are constructed through two
 distinct channels: an \textit{in situ} formation of the bulk of the
 stars in the disk and bulge of the Galaxy, followed by an \textit{ex
   situ} accumulation of substructure that builds most of the stellar
 halo~\cite{Searle:1978gc,Freeman:2002wq}. The latter stage involves
 the tidal stripping of stars and dark matter (DM) from surrounding
 satellites and subhalos and ultimately gives rise to the network of
 stellar streams that entwine the Milky
 Way~\cite{Ibata:1994fv,Ibata:2000ys,Belokurov:2006,Belokurov:2006kc,Shipp:2018yce}. The
 recent and transformative dataset of astrometry on over a billion
 stars bestowed upon us by the \Gaia
 satellite~\cite{Brown:2016tnb,GaiaDR2,Gaia:2016} has brought into
 focus the extent of this network of streams in the vicinity of
 Earth~\cite{Myeong:2017skt,Myeong:Preprint,Lancaster:2018,Malhan:2018,Meingast:2019}. This is because \Gaia
 has enabled a vast increase in the number of stars for which
 six-dimensional phase space coordinates are obtainable. While many of
 the largest and most famous stellar streams are readily identifiable
 on the sky from photometry alone~\cite{Newberg:2016}, phase space
 information is essential in identifying nearby streams that are
 dispersed over large areas of the sky.

 Here, we use a sample of \Gaia main sequence stars in the stellar
 halo, which are cross-matched with the Sloan Digital Sky Survey
 (SDSS). The chemo-dynamical
 structure of the stellar halo has been studied extensively in the last
 few years. We now know that its velocity ellipsoid exhibits two
 primary components with striking differences. One is a more metal-poor
 and essentially isotropic feature associated to the ancient, virialised
 halo. The other is more metal-rich and highly radially
 anisotropic~\cite{Ca07,My18,Be18,Ma18,Koppelman:2018}. This object was
 baptised the ``\Gaia Sausage'' after its squashed and elongated shape
 in velocity space. It is most likely the relic of a major head-on
 collision between a dwarf spheroidal and the Milky Way approximately
 8--10 billion years ago~\cite{Be18,He18,Kr18,Fattahi:2018,Vincenzo:2019}. The aspect
 of the local stellar halo that has received less detailed study however is the plethora
 of substructure~\cite{Myeong:Preprint,Myeong:2017,Myeong:2017skt,SchlaufmanECHOS1,SchlaufmanECHOS2,SchlaufmanECHOS3}. An
 extensive list of these substructures, designated ``the Shards'', 
 was presented recently in Ref.~\cite{Myeong:Preprint}, although some
 of these were anticipated in earlier searches~\cite{Smith:2009,Helmi:2017,Myeong:2017,Myeong:2017skt}. 

 The substructures are
 largely at high energy and many are members of the prominent
 retrograde tail that was known earlier~\cite{Ma12}.  The list of
 stellar substructures is surely incomplete and will continue to grow
 as further searches are made through the \Gaia data. Nevertheless, the Shards will likely 
 remain the most prominent examples locally since they have the highest significance 
 and were found in incomplete samples and with brute-force techniques~\cite{Myeong:2017,Myeong:2017skt}.
 Therefore we will use the
 Shards as typical representatives in the local stellar halo of the {\it
   ex situ} phase of the Milky Way's growth.


 Given the observations of the stellar Shards, we expect there to be
 DM counterparts: the ``Dark Shards'', relics of the DM
 subhalos that hosted the progenitors of the Shards. The material in
 streams remains kinematically cold, so the Dark Shards will have
 narrow velocity dispersions, smaller than their mean orbital
 velocities. Hence if even a small fraction of the
 local DM density is comprised of streams, the resulting speed
 distribution may be very different from the smooth Maxwellians used by
 experiments searching for DM on Earth~\cite{Lewin:1995rx, Lee:2012pf,O'Hare:2014oxa,Savage:2006qr,OHare:2017yze,Kavanagh:2016xfi}.
 Many methods for detecting DM are highly dependent on, or at the very
 least require, assumptions for the distribution of particle
 velocities in the Solar System.  High energy, incompletely phase mixed
 galactic debris accumulated relatively recently in the Milky Way's
 growth will make this distribution distinctly non-Gaussian.


 The extensive
 literature (e.g.,~Refs.~\cite{Peter:2009ak,McCabe:2010zh,Fox:2010bz,Fox:2010bu,Fox:2010bu,Peter:2011eu,McCabe:2011sr,Frandsen:2011gi,Gondolo:2012rs,Kavanagh:2013wba,Kavanagh:2013eya,DelNobile:2013cta,Fox:2014kua,Feldstein:2014gza,Gelmini:2016pei,Gondolo:2017jro,Gelmini:2017aqe,Ibarra:2017mzt,Fowlie:2017ufs,Ibarra:2018yxq,Herrero-Garcia:2019ntx,Amole:2019coq,Buch:2019aiw,Besla:2019xbx})
 on the subject of the astrophysical uncertainties of direct DM
 detection has typically only investigated the impact of changes to the
 broad shape of the speed distribution $f(v)$, often inspired by the
 variety of functional forms observed in simulations~\cite{Lisanti:2010qx,Mao:2013nda,Mao:2012hf,Butsky:2015pya}. {\it Ex situ}
 material has the potential to make more complex modifications to expected
 signals in DM experiments. While this topic has been discussed somewhat in studies
 taking inspiration from cosmological simulations~\cite{Kuhlen:2009vh,Kuhlen:2012fz}, this
 component of the local DM distribution is the most problematic for
 model-building, given that it encodes all the particularities and
 peculiarities of the formation of the Milky Way. However, the advent of
 \Gaia and the discovery of substructures like the Shards means that
 these peculiarities now have an observable counterpart. This
 data-driven perspective has been followed in several recent studies~\cite{Herzog-Arbeitman:2017fte,Herzog-Arbeitman:2017zbm,Necib_2019,Necib:2018igl,Evans:2018bqy,Ostdiek:2019gnb,Necib:2019zbk,Necib:2019zka,Bozorgnia:2019mjk}.
This work has so far dealt with the shape of $f(v)$ accounting
 for the isotropic and Sausage components observed in the \Gaia stellar
 halo.  Here, our focus is not on the overall shape, but on the
 fluctuations and deviations from Gaussianity caused by the presence of
 a host of accreted substructure. While the Dark Shards will be subdominant,
  their consequences for the velocity distribution
 are more complex and interesting than simply a shift in functional form.

 In terms of stellar content, two of the most substantial Shards are
 the S1 and S2 streams~\cite{Myeong:Preprint,Myeong:2017,Yu19}. Modelling
 suggests that their progenitors had virial masses $\approx 10^{10}
 M_\odot$ and stellar masses $\approx 10^6 M_\odot$, comparable to
 modern day dwarf spheroidal satellites like Draco and Ursa Minor.  As
 the S1 stream is counterrotating, its associated DM meets the Earth
 head-on, so it leaves characteristic signatures in experiments sensitive to high DM speeds~\cite{OHare:2018trr}. The S2 stellar
 stream is on a polar orbit that passes through the Earth. Its
 implications for DM experiments have not hitherto been studied. 
 
To study the effects of the Dark Shards on experimental DM signals in more detail, 
  the earlier sections of this paper are devoted to the refining and remodelling of the Shards from Ref.~\cite{Myeong:Preprint}.
 We begin in Sec.~\ref{sec:data} by describing our sample of halo stars
 from the SDSS-\Gaia catalogue and then discuss the stellar Shards in detail in Sec.~\ref{sec:shards}.  We then use the data and our list of substructures to inspire a model for the local DM
 distribution in Sec.~\ref{sec:darkshards}.  We discuss the new features
 present in this velocity distribution when observed in the rest frame
 of the Earth in Sec.~\ref{sec:labboost}.  In the remaining two sections,
 we detail the consequences of these new features in experiments
 looking for axions and axion-like particles (Sec.~\ref{sec:axions}), and nuclear recoil-based searches for DM
 (Sec.~\ref{sec:recoils}). In the latter case we also discuss the effects on annual modulation
 and directional signals.  We conclude in Sec.~\ref{sec:summary}. 
 
 A GitHub repository~\cite{DarkShards}
is available which contains links to {\sc IPython} notebooks for 
individually reproducing each result.

 \section{The structure of the local stellar halo}\label{sec:data}

 \begin{figure*}[!t]
 \includegraphics[width=0.75\textwidth,trim={0cm 0cm 0cm 0cm},clip]{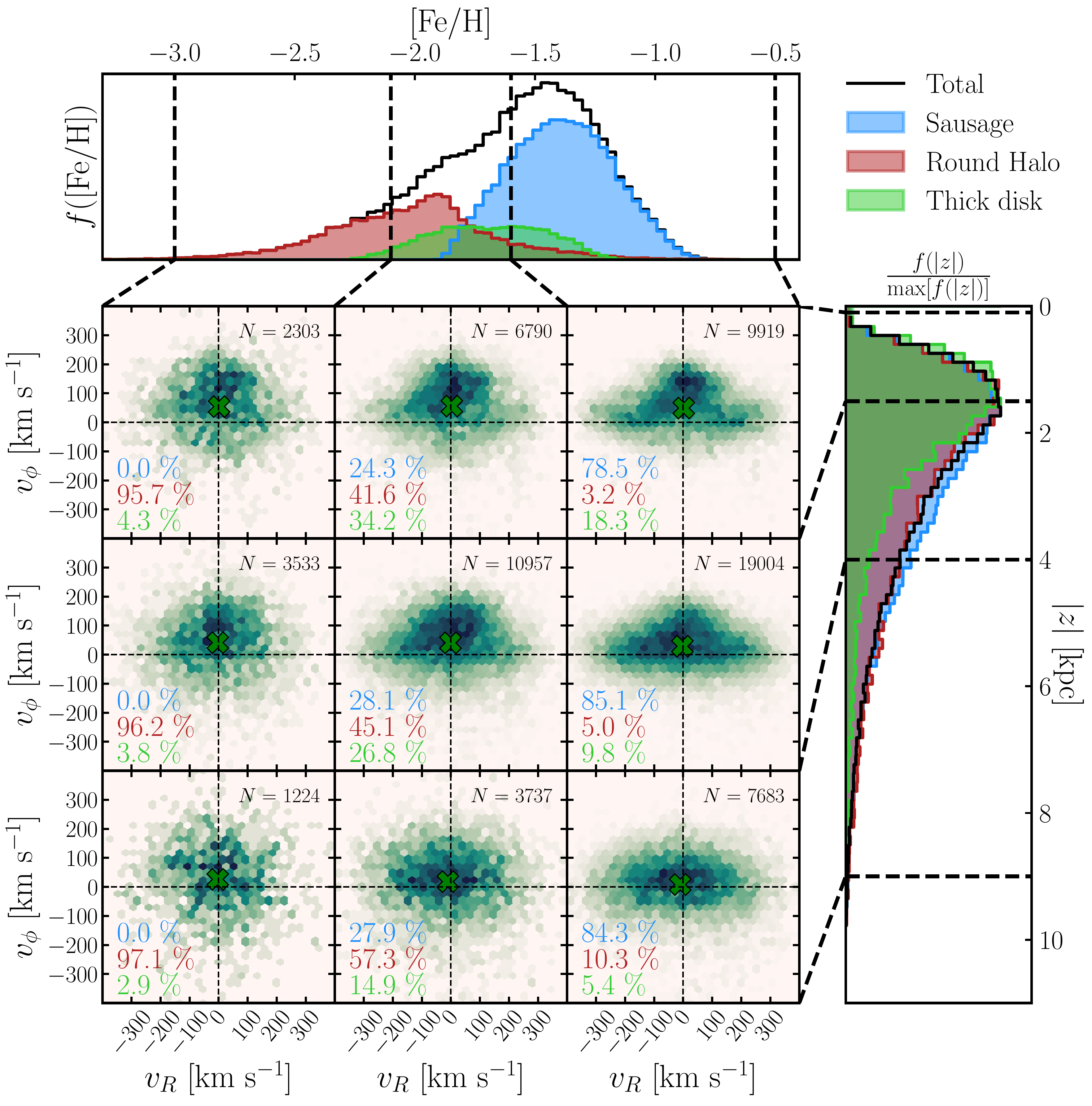}
 \caption{Kinematic structure of the stellar halo sample. The central
   panels show 2-d histograms of the radial and azimuthal velocities,
   $v_R$ and $v_\phi$, of the SDSS-\Gaia stellar halo sample used in
   this paper.  Darker colours represent a higher star count.  The
   full sample has been divided into nine bins in terms of metallicity
   ($[{\rm Fe/H}]$) and distance away from the disk plane ($|z|$), as
   indicated by the dashed black lines (following Ref~\cite{Be18}).
   Reading from left to right corresponds to increasing $[{\rm
       Fe/H}]$, while from top to bottom corresponds to increasing
   $|z|$.  The green cross shows the mean velocity in each bin, while
   $N$ gives the number of stars (there are 261 stars with $[{\rm
       Fe/H}]<-3$ or $|z|>9$ that are not included in the 2-d
   histograms). We decompose the sample into three components: Round
   Halo, Sausage and Thick disk. Their 1-d $[{\rm Fe/H}]$ and~$|z|$
   distributions are shown above and to the right of the
   histograms. The fraction of each component in the nine bins is
   given by the coloured percentage values.}
 \label{fig:vrvphi}
 \end{figure*}

 The SDSS-\Gaia catalogue enhances and recalibrates the astrometric
 solution of the SDSS's 9th data release~\cite{Ahn:2012fh} with
 parallaxes and proper motions from \Gaia~\cite{Gaia:2016,Brown:2016tnb}. 
 This catalogue expands our view of
 the Milky Way halo out to heliocentric distances of approximately
 10~kpc. The majority of the radial velocities are sourced from SDSS; however a minor fraction are cross-matched with
 spectroscopic surveys APOGEE~\cite{Anders:2013gna}, LAMOST
 DR3~\cite{Luo15} and RAVE-on~\cite{Casey:2017,RAVE5}. Since photometric parallaxes are calculable for main
 sequence turnoff~\cite{Ivezic:2008wk} and blue horizontal branch
 stars~\cite{Deason:2011wt}, we use these stars to construct a
 sample of the disk and halo with six-dimensional phase space information
 out to approximately 10~kpc.

 To extract the stellar halo from this selection of stars, we adopt a
 similar set of cuts to those in
 Refs.~\cite{Myeong:2017skt,Myeong:Preprint}.  To ensure a high quality
 sample, stars are selected with photometric parallax distances
 \mbox{$<10$~kpc}, a heliocentric radial velocity error \mbox{$<15
   \kms$}, and distance errors \mbox{$<2.5$~kpc}. 
    Then, and importantly
 for this work, stars with clear membership to the galactic thin and
 thick disks have been removed. Disk stars occupy positive azimuthal
 velocities, $v_\phi \gtrsim 100 \kms$ and high metallicities $[{\rm
     Fe/H}] \gtrsim -1$ (see e.g.,\ Fig.~1 of
 Ref.~\cite{Myeong:2017skt}). The cut of disk stars is based on these two parameters, but to leave as much of the halo intact as possible it has not been applied too aggressively. Therefore some contamination remains from the metal-poor tail of the thick disk which overlaps strongest with the halo in the [Fe/H]-$v_\phi$ plane. We will see this contamination for metallicities $[{\rm Fe/H}]\sim -2.1$--$-1.6$. After the cuts, the sample contains $65406$ stars.

 \subsection{The smooth component}

 Figure~\ref{fig:vrvphi} shows 2-d histograms of the stellar halo sample
 in~$v_R$ and~$v_\phi$, the radial and azimuthal
 velocities.\footnote{Throughout, we use ($R,\phi, z$) as
   galactocentric cylindrical polar coordinates and ($r,\theta,\phi$)
 as galactocentric spherical polars.}
 We divide the total sample into nine histograms, with cuts on the
 height above or below the disk~$|z|$ and iron abundance~[Fe/H], as
 indicated by the black dashed lines above and to the right of the nine
 histograms.  A striking observation can be made when the metal rich
 panels (right) are compared with the metal-poor panels (left):
 the distribution of stars is clearly extended in $v_R$ in the
 right-most panels, making it sausage-shaped.  In contrast, the
 distribution of stars in the left-most panels is circular, consistent
 with a roundish, isotropic halo.  We will refer to this as the ``Round Halo'' to distinguish it from the
 radially anisotropic ``Sausage''.  Some contamination by thick disk
 stars can also be seen in the top-row middle panel at $v_\phi\simeq 125 \kms$ and $v_R\simeq0.0
 \kms$.

 Above and to the right of the 2-d histograms, we show the 1-d histograms of $|z|$
  and [Fe/H] for the three components: the Round Halo (red),
 Sausage (blue) and thick disk (green). To partition the stars in this
 way, we have decomposed the sample by applying a Gaussian mixture
 model using the stellar velocity and metallicity data: $\mathbf{q} =
 \{v_R,v_\phi,v_z,[{\rm Fe/H}]\}$. The probability density function for
 the three components that we fit to the stellar sample is,
 \begin{equation}\label{eq:gaussian}
 \begin{split}
 f(\mathbf{q}) &= \sum_{i=1}^{3} \frac{w_i}{\sqrt{(2\pi)^D \det{\boldsymbol{\Sigma}_i}}} \\
 &\qquad \quad \times \exp\left(-\frac{1}{2} (\mathbf{q}-\bar{\mathbf{q}}_i)^T \,\boldsymbol{\Sigma}_i^{-1}\, (\mathbf{q}-\bar{\mathbf{q}}_i)\right) \, ,
 \end{split}
 \end{equation}
 where $\boldsymbol{\Sigma}_i$ is a covariance matrix for
 component~$i$, $w_i$ is its relative weight, and the exponent $D$ is
 the dimensionality of the data, which in this case is four.  In
 Table~\ref{tab:fit}, we give the means, weights and covariances for
 this decomposition for the complete stellar halo sample.

 As has been noted before~\cite{My18,Be18,Ma18,Haywood:2018}, the
 Sausage component dominates the local stellar halo in main sequence
 stars. It has also been identified independently with other tracers
 including blue horizontal branch
 stars~\cite{Lancaster:2019,Deason:2011wt},
 RR~Lyraes~\cite{Watkins:2009im,He18} and K
 giants~\cite{Bird:2019}. The characteristic range of metallicities,
 [Fe/H]$ \approx -1.4$, means that its stars are slightly more
 metal-rich than the halo average ($\approx-1.9$) but metal-poor
 relative to the (thin and thick) disks ($\approx-0.8$). This strongly
 suggests that the Sausage component is the remnant of a merger in the
 Milky Way's recent past. The unusual kinematic profile implies that
 the merger was a head-on collision with an object on a low inclination
 orbit~\cite{Be18}.  The progenitor Sausage galaxy most likely had a
 total mass $\gtrsim 10^{10}\, M_\odot$ and was accreted around
 redshift $\lesssim 3$~\cite{Be18,Fattahi:2018,Mackereth:2019}.

 \begin{table}[t!]
 \ra{1.3}
 \begin{tabularx}{0.49\textwidth}{X|XXX}
 \hline\hline
       & {\bf Round Halo}& {\bf Sausage} & {\bf Thick disk}  \\
       \hline
 $\bar{v}_r[{\rm km\,s}^{-1}]$ & $-0.1$  & $-8.2$  &  $12.4$ \\
 $\bar{v}_\phi[{\rm km\,s}^{-1}]$ & $6.0$ & $25.7$ & $128.8$ \\
 $\bar{v}_z[{\rm km\,s}^{-1}]$ & $8.0$ & $0.99$ & $3.6$ \\
 $\sigma_r[{\rm km\,s}^{-1}]$ & $144.4$  &  $158.9$  & $76.8$ \\
 $\sigma_\phi[{\rm km\,s}^{-1}]$ & $120.0$ & $61.5$ & $50.1$ \\
 $\sigma_z[{\rm km\,s}^{-1}]$ & $115.7$ & $80.9$ & $62.7$ \\
 $\Sigma_{r\phi}/\sigma_r\sigma_\phi$ & $0$ & $0$ & $0$ \\
 $\Sigma_{rz}/\sigma_r\sigma_z$ & $0.1$ & $0.3$ & $0.2$ \\
 $\Sigma_{\phi z}/\sigma_\phi\sigma_z$ & $0$ & $0$ & $0$ \\
 $\overline{\rm [Fe/H]}$ & $-1.9$ & $-1.4$ & $-1.7$ \\
 $\sigma_{\rm [Fe/H]}$ & $0.4$ & $0.2$ & $0.3$ \\
 $w$ & $0.33$ & $0.51$ & $0.16$ \\
        \hline\hline
 \end{tabularx}
 \caption{Results of a three population Gaussian mixture model fit to
   the SDSS-\Gaia stellar halo sample. In order, we show the mean velocities, velocity dispersions (diagonal covariance matrix elements),
   the off-diagonal covariance matrix elements for the velocities, mean metallicity, standard deviation on the metallicity, and finally the
    weighting of each component $w$. The thick disk is included as
   a third population to model its small contamination of the sample.}
 \label{tab:fit}
 \end{table}

 \subsection{The Shards}
 \begin{figure*}[!t]
 \includegraphics[width=0.99\textwidth,trim={0cm 0cm 0cm 0cm},clip]{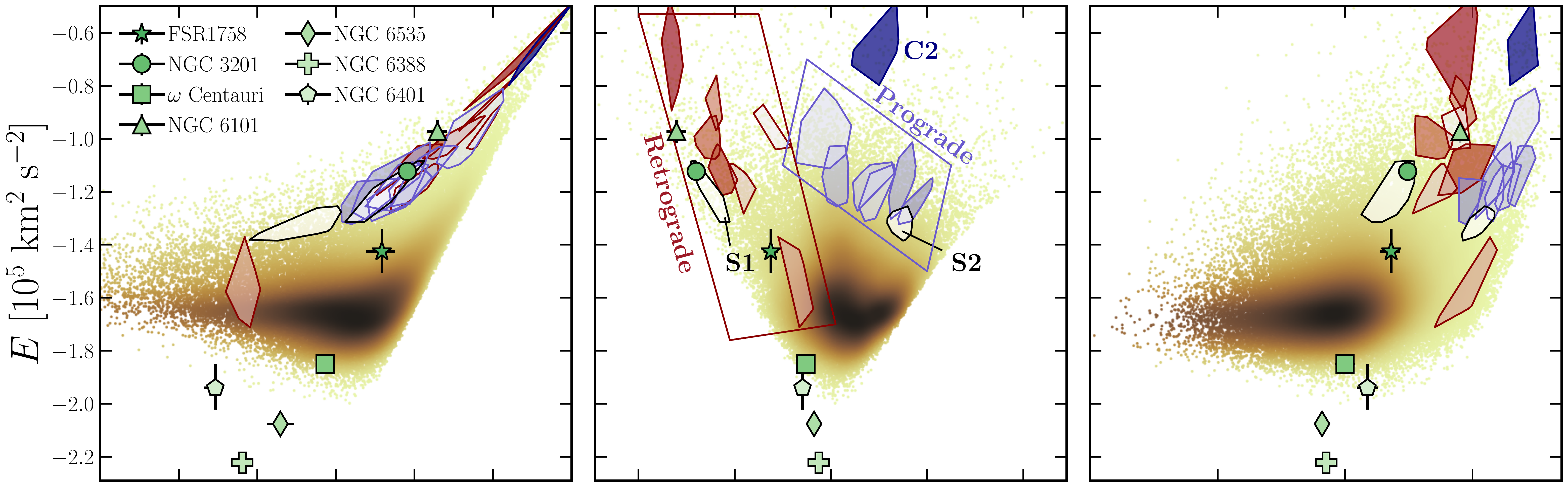}
 \includegraphics[width=0.99\textwidth,trim={0cm 0cm 0cm 0cm},clip]{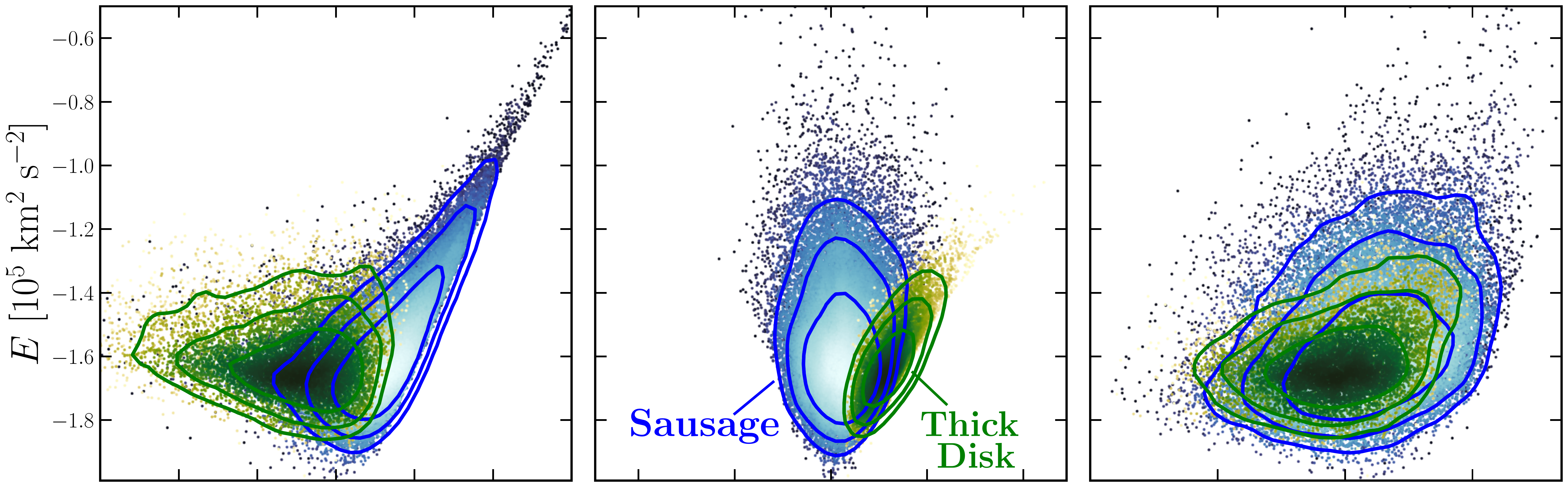}
 \includegraphics[width=0.99\textwidth,trim={0cm 0cm 0cm 0cm},clip]{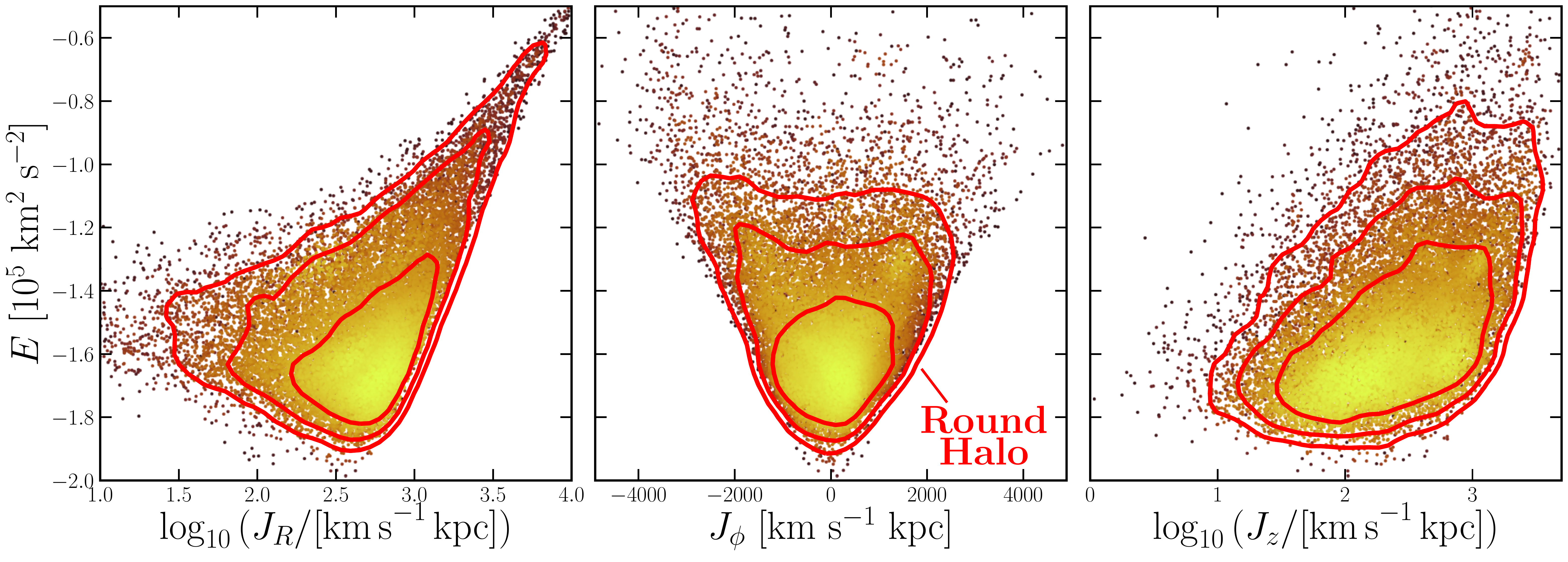}
 \caption{Action-energy distributions for the full sample of stars (top
   row) and after being partitioned into the Sausage and Thick Disk
   (middle row), and Round Halo (bottom row). In the top row, we show
   the convex hulls of stars belonging to the Shards. We highlight the
   three highest significance Shards (S1, S2 and C2) and group the
   retrograde and prograde candidate substructures together. 
   The middle and bottom rows show subsets of the full
   sample that are partitioned according to the decomposition
   used in Fig.~\ref{fig:vrvphi} and summarised in
   Table~\ref{tab:fit}. We show logarithmically
   spaced contours over the distributions to highlight the general
   shapes, for example the high energy retrograde tail of negative $J_\phi$ in the
   bottom row. In the top row we also
   show the seven retrograde globular clusters that are potentially
   linked with the Sequoia event producing S1 and the retrograde tail. 
   }
 \label{fig:actions_decomposed}
 \end{figure*}

 The Round Halo and the Sausage are apparent when visualising the
 full sample in velocity space. The importance of the {\it ex-situ}
 halo, the accreted part of our Galaxy, on the other hand is more evident in action
 space~\cite{Binney:2012wv}. Actions are adiabatic invariants so are
 conserved under slow evolution of the potential. This means that stars
 belonging to a discrete bound object that are accreted together slowly
 will remain clustered in action space even if they are too dispersed
 or too close to us to be identified otherwise. Searching through
 action space is therefore the most efficient and powerful way to find
 local substructure. In addition to the three actions
 $(J_R,J_\phi,J_z)$, another useful quantity is the orbital energy,
 $E$, as substructure from recent in-falls is usually at high energy.
 The distribution of the full sample in these three actions versus
 energy are shown in the top row of
 Fig.~\ref{fig:actions_decomposed}. Actions are computed using the
 numerical methods reviewed in Ref.~\cite{Sanders2016} and the updated
 Milky Way potential from McMillan~\cite{McMillan:2017}.

 The action space distribution of the sample contains abundant substructure which is highlighted in the top row of Fig.~\ref{fig:actions_decomposed}. The 30 highest significance clusters
 were initially listed in Ref.~\cite{Myeong:Preprint}; however the
 list in fact extends to a total of 69. We call all these action-space clusters
 ``Shards'' throughout our paper. Action space clusterings are typically associated with kinematically cold streams, however only a fraction of the reported Shards are significant enough for this interpretation to be ascribed. Those action space clusters that are not focused tightly around single streaming velocities are sometimes referred to as ``clumps'', ``objects'' or ``moving groups''~\cite{Myeong:2017skt}. For example our list of Shards include the prominent S1 and S2 streams as well as the clump C2, themselves rediscoveries of earlier reported objects found
 in phase space~\cite{Myeong:2017,Myeong:2017skt}. For clarity, we refer to S1 and S2 as ``streams'', and simply use the term ``substructures'' for the remaining Shards. In Fig.~\ref{fig:actions_decomposed} we grouped them together into two
 categories ``Retrograde'' and ``Prograde'' according to the sign of their
 $J_\phi$ (these correspond to the Rg and Cand labels used in
 Ref.~\cite{Myeong:2017skt}). In total, the Shards comprise a subset of
 1117 stars in the \Gaia-SDSS halo sample.\footnote{While the Shards were originally identified in the \Gaia-SDSS sample with \Gaia-DR1 proper motions, we have verified that the fitted parameters of the S1 and S2 streams are essentially unchanged when the slightly improved \Gaia-DR2 is used. This is likely to be the case for the majority of the Shards.}

 The Retrograde Shards contain some of the highest energy stars in the
 halo and seem to be comparatively metal-rich, with $-1.9\lesssim~[{\rm
     Fe/H}]~\lesssim~-1.3$.  While a similar range of 
 metallicities is found for the Sausage stars, the characteristic abundances of
 other $\alpha$-elements in the retrograde tail clearly distinguishes
 them in origin from the Sausage
 galaxy~\cite{Stephens:2002,Venn:2004,Matsuno:2019}. It has been
 suggested that this retrograde tail and the substructures it contains
 may all be associated with another major low-inclination accretion of
 a $10^{10}M_\odot$ dwarf galaxy, dubbed the
 ``Sequoia''~\cite{Myeong:2019a}. This interpretation is bolstered by
 several globular clusters on retrograde orbits that we have have also
 marked on Fig.~\ref{fig:actions_decomposed}. These include the
 extended globular cluster FSR1758~\cite{Barba:2019,Simpson:2019}
 ---from which the arboreal moniker originates----as well as the anomalous
 $\omega$ Centauri, which has for many years been believed to have begun life as the
 nucleus of a dwarf galaxy~\cite{Bekki:2003qw}. These clusters may have
 all originally been part of the Sequoia galaxy. One of the most
 important Shards, S1---already the subject of some discussion in the
 literature~\cite{OHare:2018trr,Knirck:2018knd,Buckley:2019skk}---
 appears to be associated with the globular cluster NGC 3201 which
 may have resided within the Sequoia.

 \section{Modelling the Shards}\label{sec:shards}

 \begin{figure*}[!t]
  \includegraphics[width=0.49\textwidth,trim={0cm 0cm 0cm 0cm},clip]{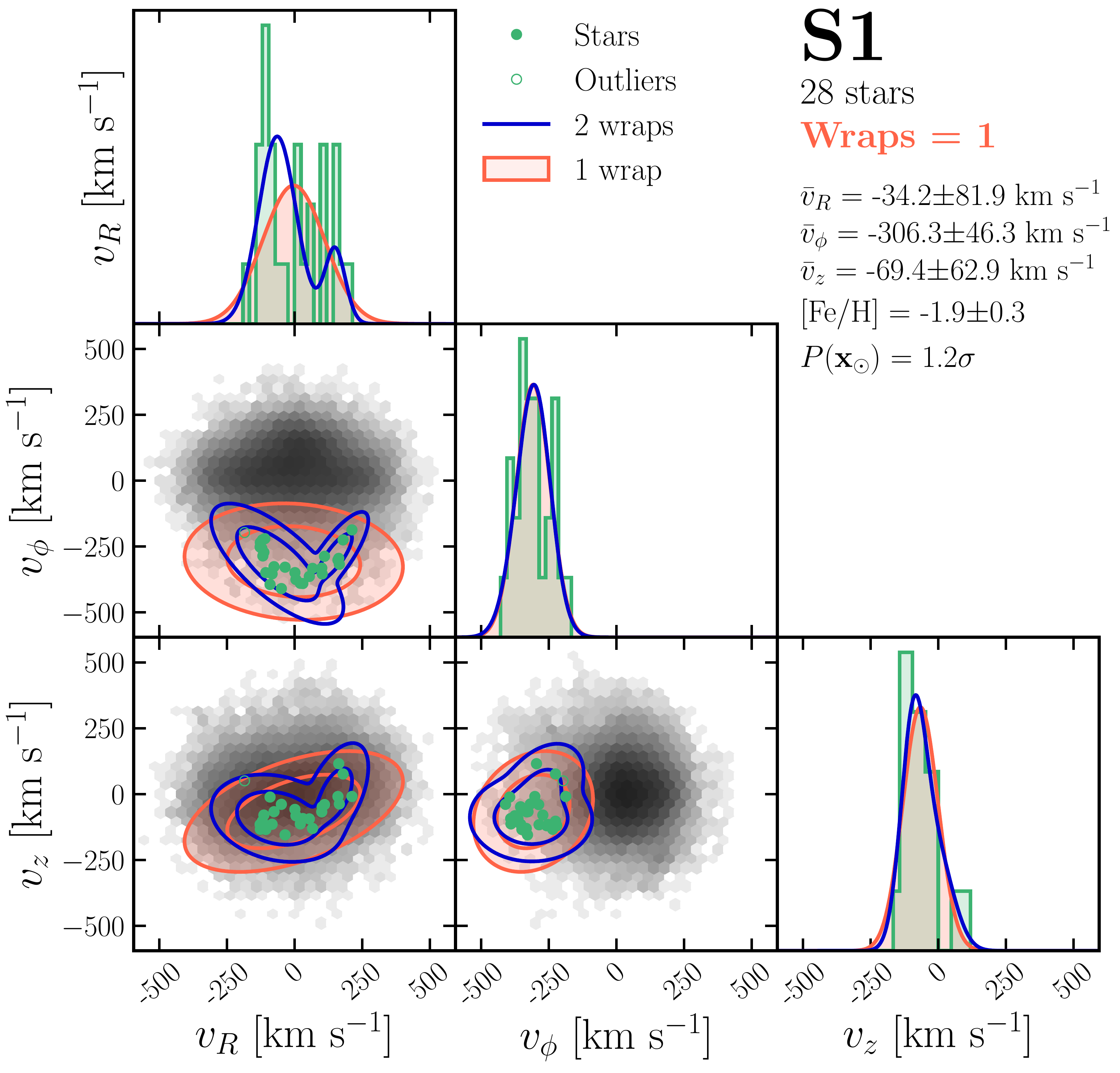}
  \includegraphics[width=0.49\textwidth,trim={0cm 0cm 0cm 0cm},clip]{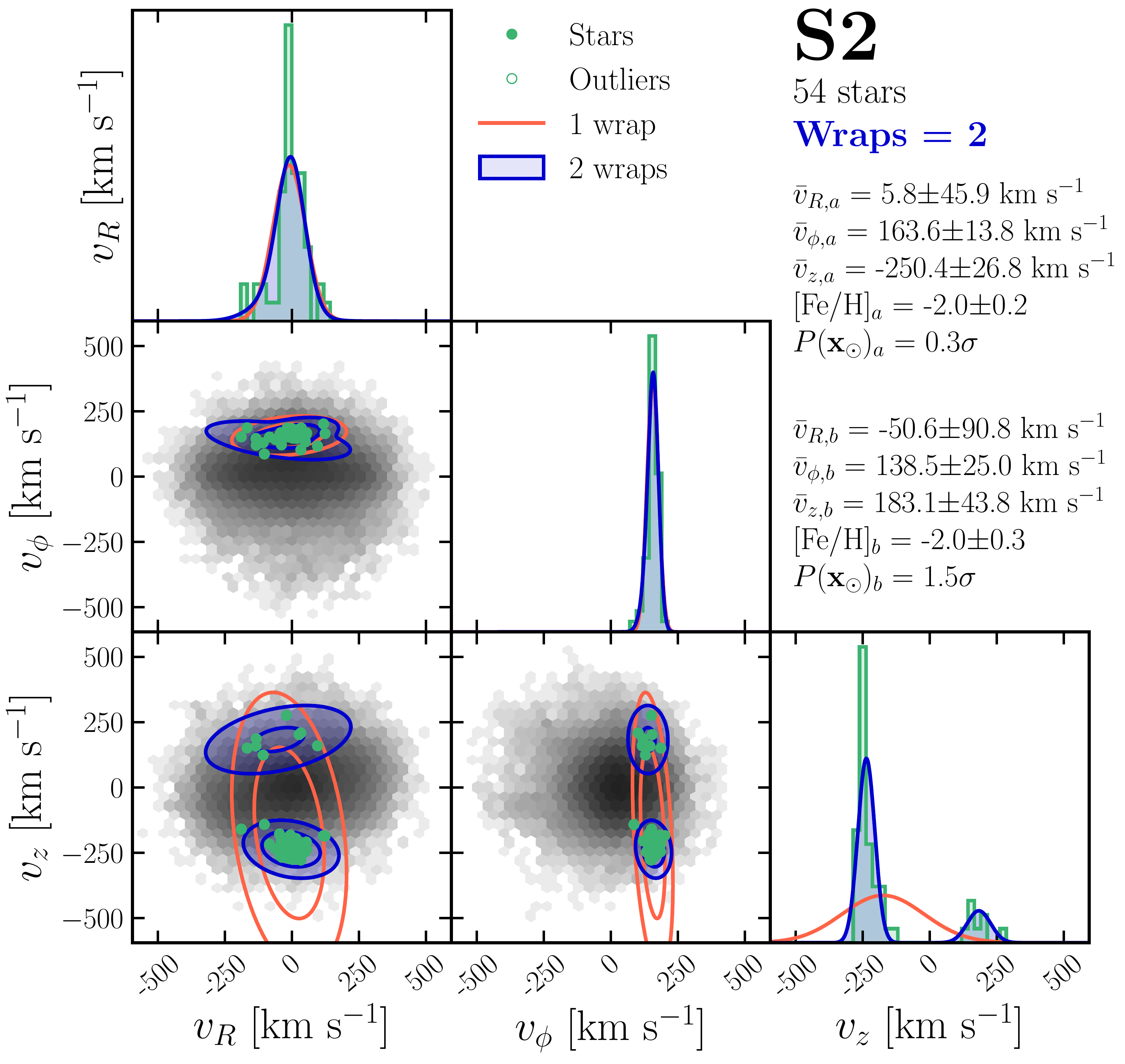}
  \includegraphics[width=0.49\textwidth,trim={0cm 0cm 0cm 0cm},clip]{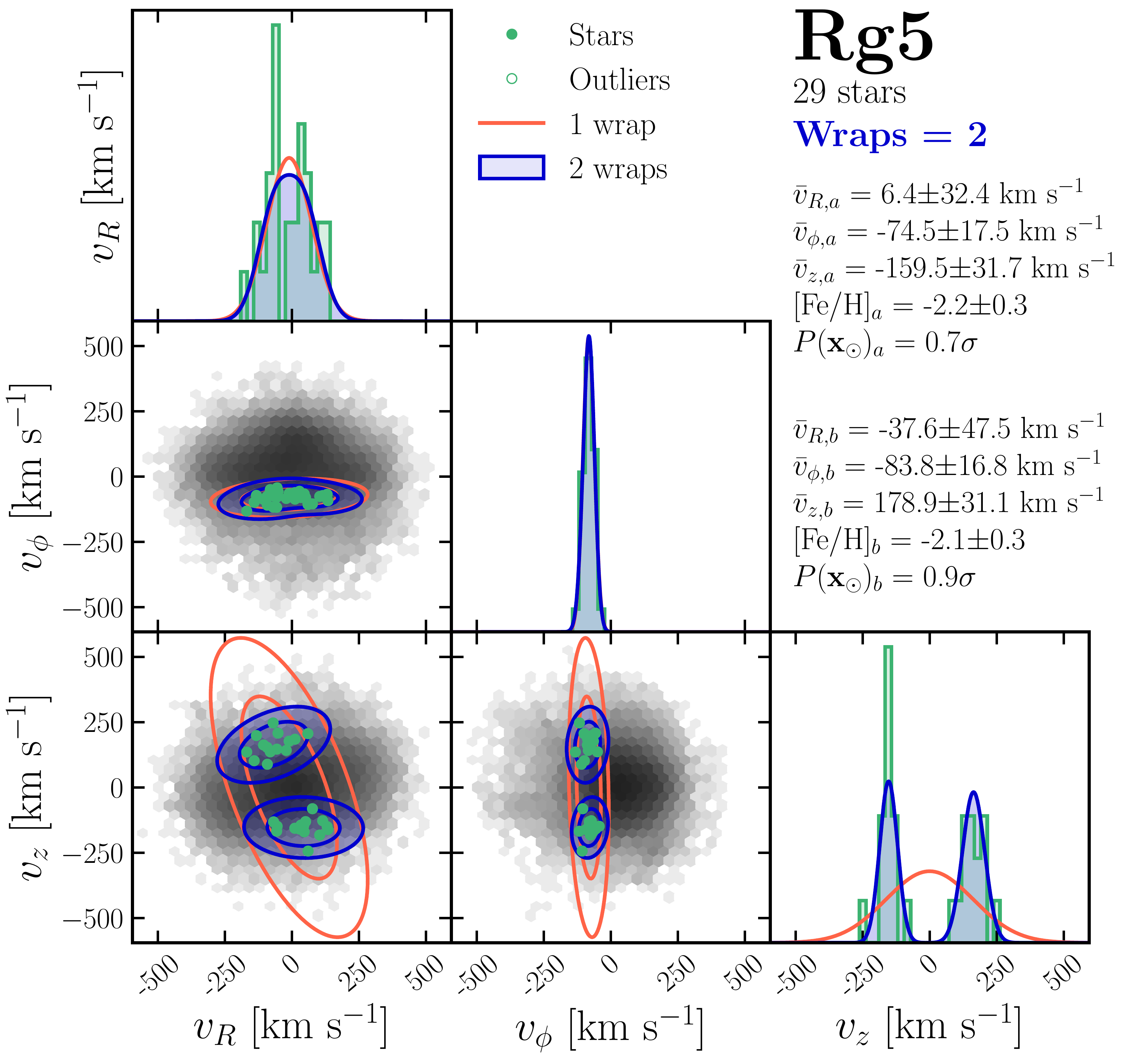}
  \includegraphics[width=0.49\textwidth,trim={0cm 0cm 0cm 0cm},clip]{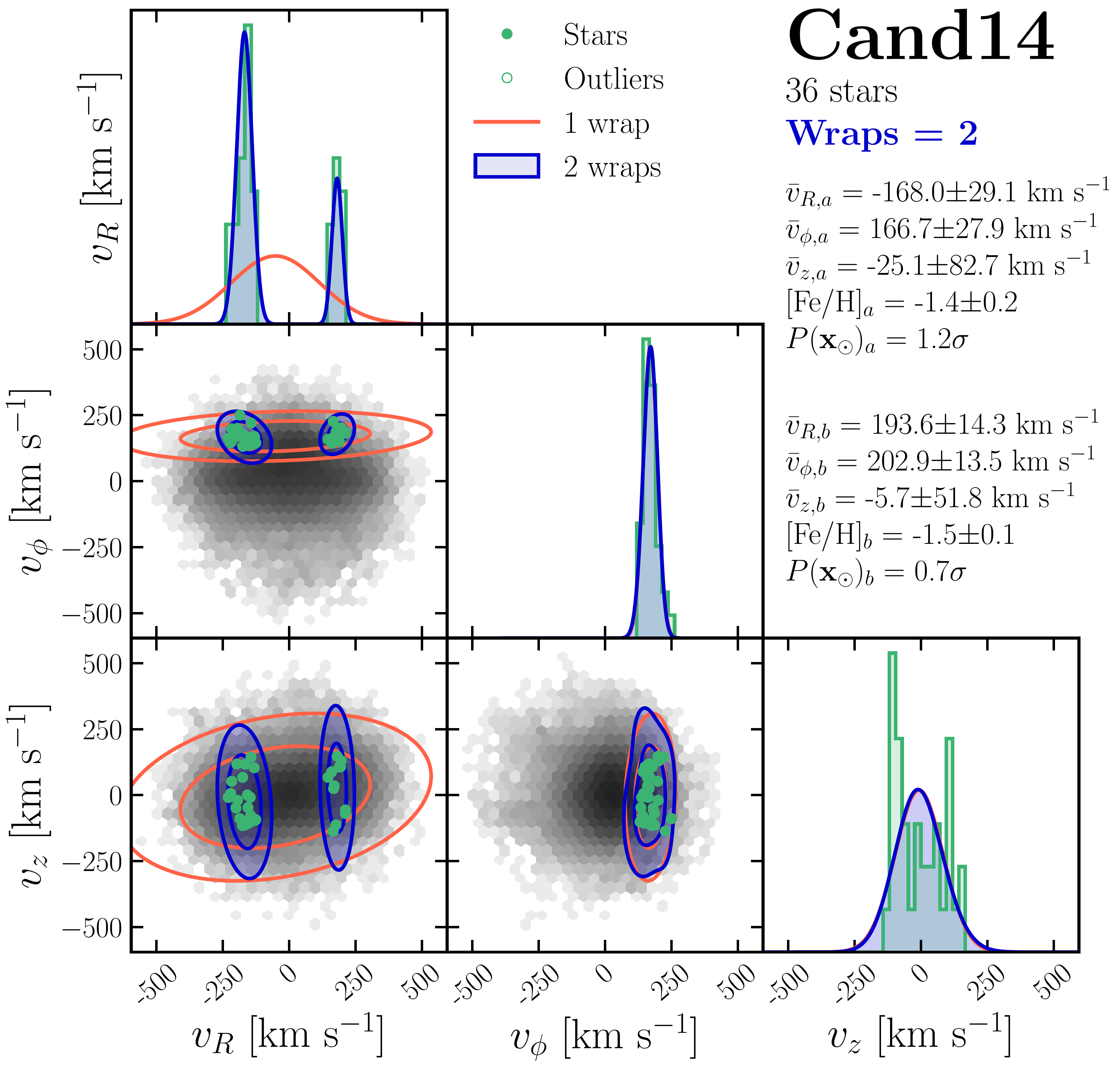}
 \caption{Distributions of stellar velocities in galactic cylindrical
   coordinates for four Shards: S1, S2, Rg5 and Cand14.  The velocity
   of stars associated with each Shard are shown as green points.  Fits
   to the seven-dimensional (position, velocity and metallicity) data with
   the Gaussian mixture model [Eq.\eqref{eq:gaussian}] are shown: the
   orange contour indicates an $n=1$ Gaussian whereas the blue contours
   show the $n=2$ case.  The extent of the SDSS-\Gaia sample is shown
   in grey.  For each Shard, we select the number of wraps
   by using the minimum Bayesian information criterion.  S1 is
   better fit by one wrap, while S2, Rg5 and Cand14 are better fit by
   two wraps.  The best-fit parameters are given in the accompanying
   text. The significance value $P(\textbf{x}_\odot)$ quantifies how far away the solar position is relative to the spatial distribution of each Shard.
   }
 \label{fig:Vtriangle}
 \end{figure*}

 \begin{figure*}
  \includegraphics[width=0.78\textwidth,trim={0cm 0cm 0cm 0cm},clip]{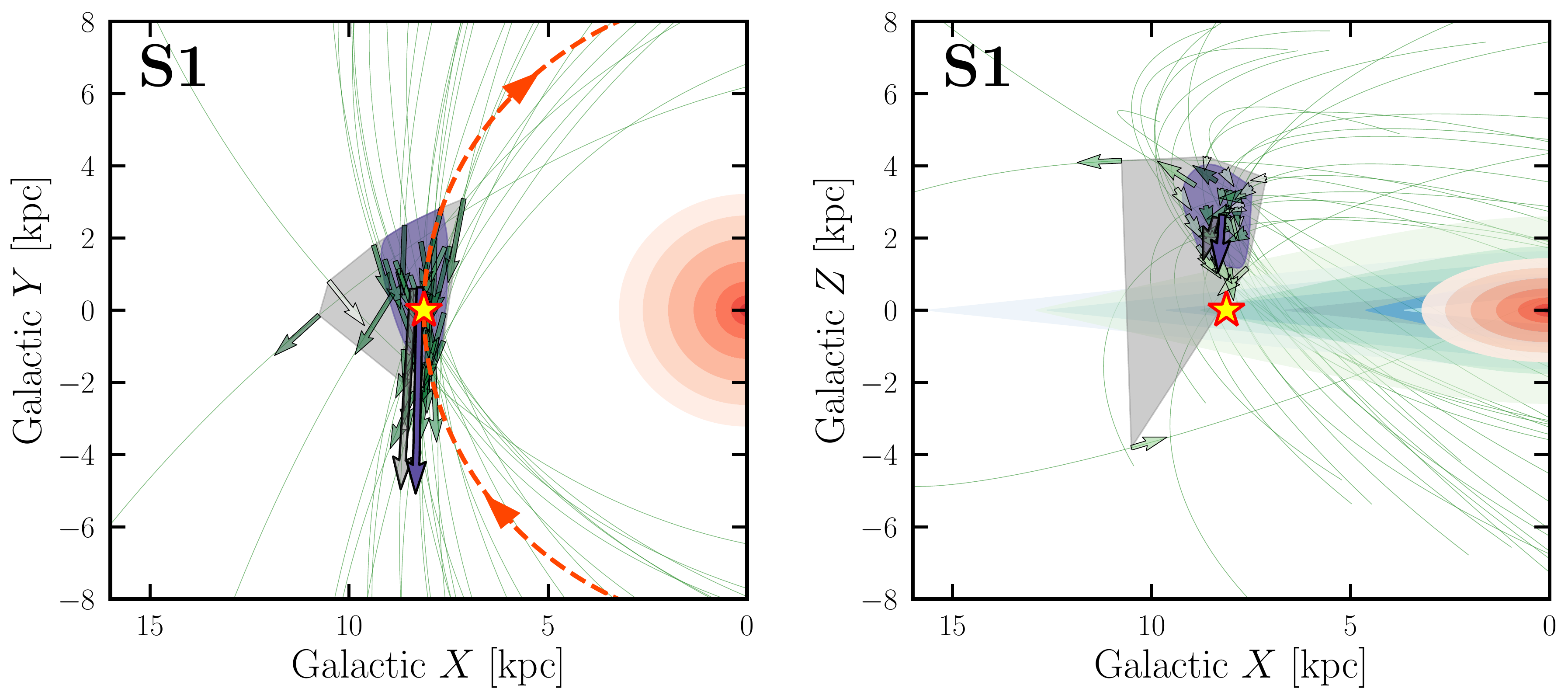}
  \includegraphics[width=0.78\textwidth,trim={0cm 0cm 0cm 0cm},clip]{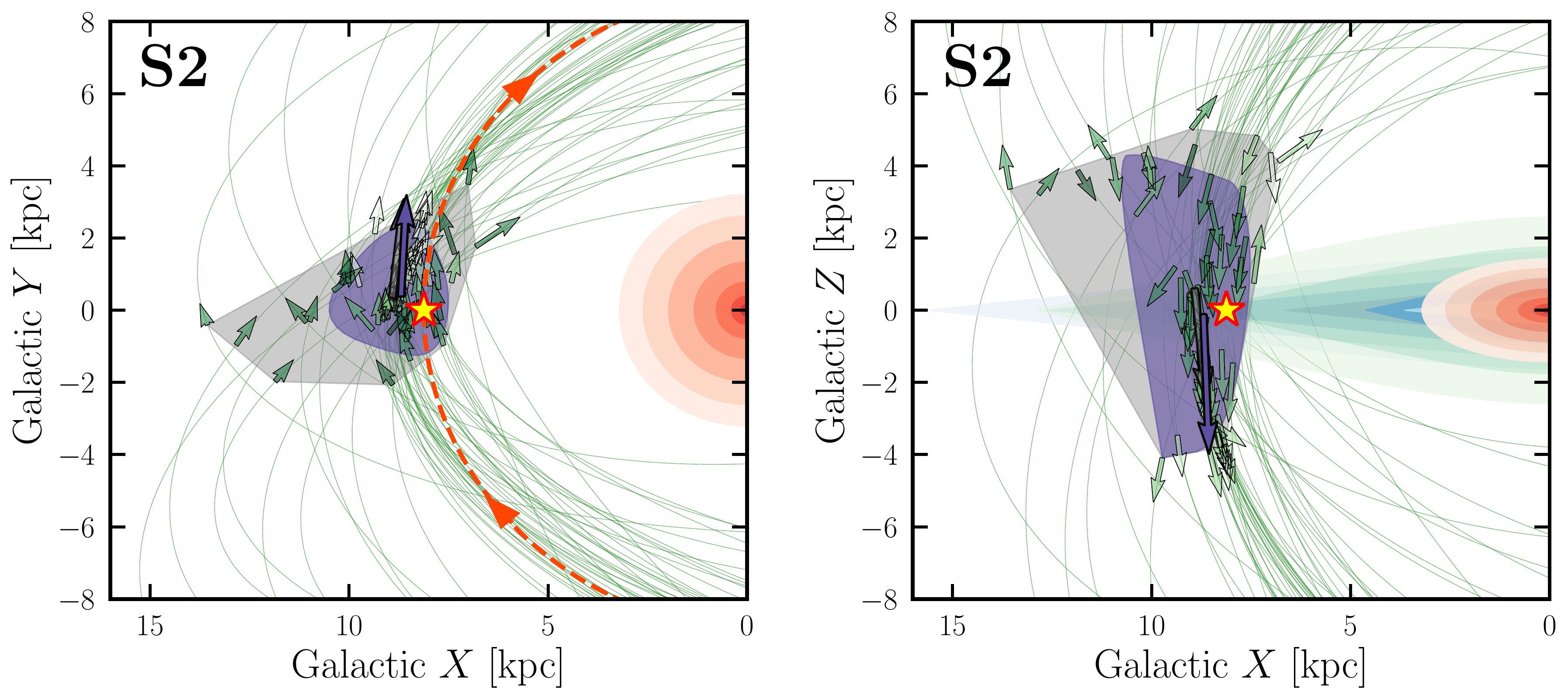}
  \includegraphics[width=0.78\textwidth,trim={0cm 0cm 0cm 0cm},clip]{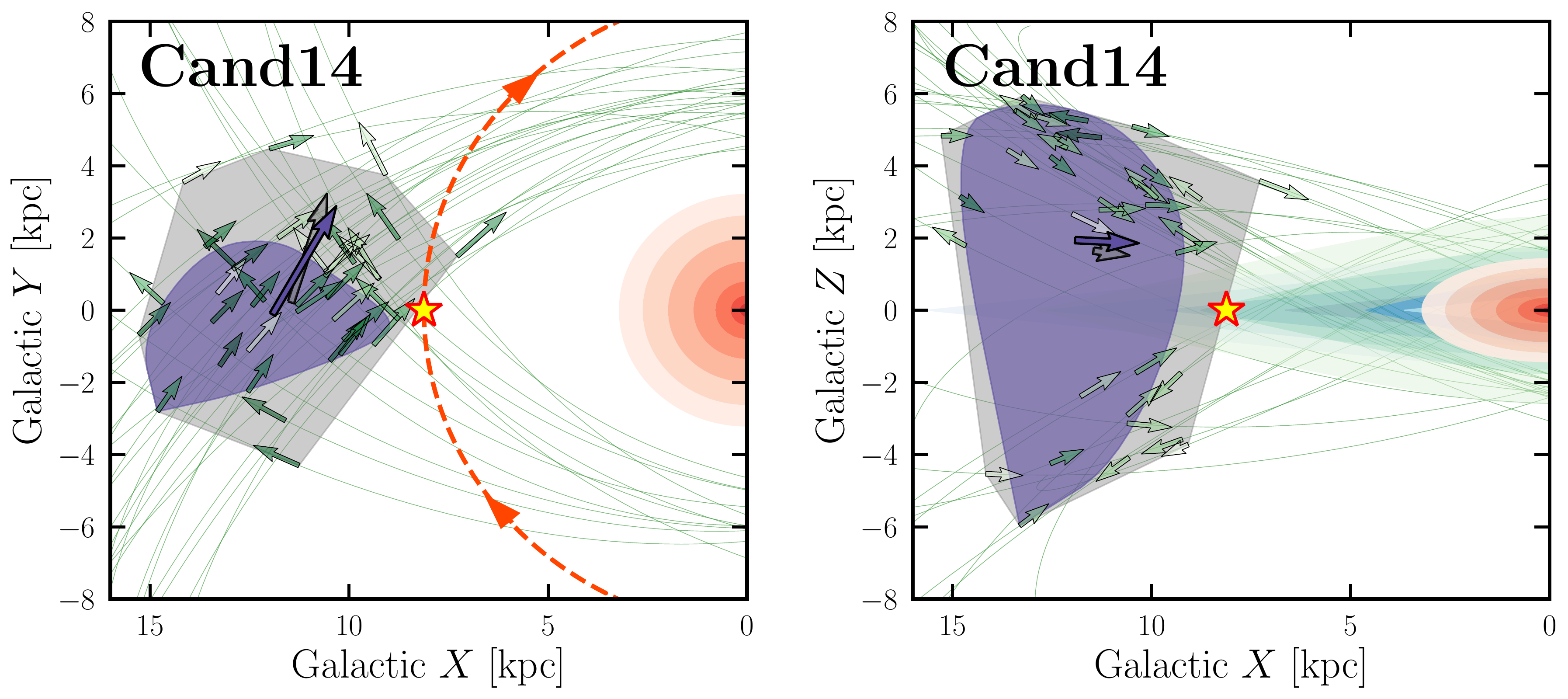}
 \caption{Stellar positions, velocities and orbits in the galactic
   $X$-$Y$ and $X$-$Z$ planes for three Shards: S1, S2 and Cand14 (from
   top to bottom). The orange region traces the shape of the density
   profile of the galactic bulge whereas the greenish and blueish
   regions correspond to the thin and thick disk respectively. The
   convex hull for all the stellar positions is shown as the outer gray
   shaded region, within which we show a reduced region after
   performing a $z$-score cut of spatial outliers. The orbit and position of the Sun is
   shown as an orange-red dashed line and yellow star. The orbits of the each star integrated forward and backward in time by 100 Myr are shown as green lines.
   }
 \label{fig:XYZ}
 \end{figure*}

 \begin{table*}[t!] 
\ra{1.3}
\begin{tabularx}{0.98\textwidth}{cc|ccccccc}
\hline\hline
 \multicolumn{2}{c|}{{\bf Name}}       & Number & $(X,Y,Z)$ & $(\Delta X,\Delta Y,\Delta Z)$ & $(v_R,v_\phi,v_z)$ & $(\sigma_R,\sigma_\phi,\sigma_z) $ & \quad $\langle [{\rm Fe}/{\rm H}] \rangle$ \quad        & \quad$P(\mathbf{x}_\odot)$\\
     &  &  of stars & kpc & kpc \quad\quad& $\kms$   \quad    & $\kms $ \quad &   &    $(\sigma)$   \\
\hline\hline  \parbox[t]{5mm}{\multirow{1}{*}{\rotatebox[origin=c]{90}{{\bf S1}}}} &        & 28        & $(8.4,0.6,2.6)$   & $(0.7,1.8,2.2)$                      & $(-34.2,-306.3,-64.4)$   & $(81.9,46.3,62.9)$                & $-1.9\pm 0.3$ & 1.2 \\
\hline\hline  \parbox[t]{5mm}{\multirow{2}{*}{\rotatebox[origin=c]{90}{{\bf S2}}}} & a      & 46        & $(8.7,0.4,0.1)$   & $(0.7,1.2,6.9)$                      & $(5.8,163.6,-250.4)$     & $(45.9,13.8,26.8)$                & $-2.0\pm 0.2$ & 0.3 \\
& b      & 8        & $(10.1,0.2,3.3)$   & $(4.9,0.7,1.4)$                      & $(-50.6,138.5,183.1)$    & $(90.8,25.0,43.8)$                & $-2.0\pm 0.3$ & 1.5 \\
\hline\hline \parbox[t]{5mm}{\multirow{7}{*}{\rotatebox[origin=c]{90}{{\bf Retrograde}}}} & Rg2      & 13        & $(8.9,0.3,4.4)$   & $(0.8,2.1,2.7)$                      & $(44.5,-248.4,185.2)$    & $(105.9,23.1,63.5)$               & $-1.6\pm 0.2$ & 1.4 \\
& Rg5a     & 15        & $(8.4,0.8,1.1)$   & $(1.0,1.3,3.3)$                      & $(6.4,-74.5,-159.5)$     & $(32.4,17.5,31.7)$                & $-2.2\pm 0.3$ & 0.7 \\
& Rg5b     & 14        & $(8.1,-0.2,2.2)$  & $(1.1,1.2,2.4)$                      & $(-37.6,-83.8,178.1)$    & $(47.5,16.8,31.1)$                & $-2.1\pm 0.3$ & 0.9 \\
& Rg6a     & 17        & $(8.3,0.2,3.3)$   & $(1.8,1.4,2.0)$                      & $(105.1,-230.2,202.4)$   & $(73.7,16.8,86.6)$                & $-1.6\pm 0.2$ & 1.1 \\
& Rg6b     & 12        & $(8.5,0.9,3.2)$   & $(1.5,1.5,2.2)$                      & $(-233.2,-221.8,51.6)$   & $(32.7,14.4,115.7)$               & $-1.7\pm 0.3$ & 0.6 \\
& Rg7a     & 5         & $(8.2,0.5,3.3)$   & $(2.1,1.5,3.3)$                      & $(309.0,-191.3,-83.4)$   & $(66.7,17.1,102.7)$               & $-1.5\pm 0.1$ & 1.1 \\
& Rg7b     & 9         & $(8.9,-0.0,5.1)$  & $(1.9,1.3,2.0)$                      & $(-288.7,-158.1,-105.5)$ & $(78.7,65.8,111.8)$               & $-1.5\pm 0.3$ & 1.8 \\
\hline\hline \parbox[t]{5mm}{\multirow{16}{*}{\rotatebox[origin=c]{90}{{\bf Prograde}}}} & Cand8a   & 31        & $(9.9,-0.1,2.4)$  & $(2.1,2.5,4.4)$                      & $(-6.7,207.7,-186.4)$    & $(114.6,20.8,73.5)$               & $-1.8\pm 0.4$ & 0.4 \\
& Cand8b   & 18        & $(8.4,0.6,1.1)$   & $(1.5,2.2,3.6)$                      & $(33.6,213.9,214.1)$     & $(96.5,22.7,37.7)$                & $-1.8\pm 0.2$ & 0.1 \\
& Cand9    & 43        & $(9.2,-0.2,1.7)$  & $(1.1,1.4,3.4)$                      & $(11.0,177.5,-251.4)$    & $(120.6,13.9,132.2)$              & $-1.8\pm 0.2$ & 0.5 \\
& Cand10   & 38        & $(8.6,-0.0,2.0)$  & $(1.7,1.3,2.5)$                      & $(-37.4,20.0,192.3)$     & $(161.5,18.2,195.0)$              & $-2.0\pm 0.2$ & 0.2 \\
& Cand11a  & 14        & $(9.1,-0.3,2.7)$  & $(2.5,1.4,3.8)$                      & $(36.8,116.5,-271.5)$    & $(96.1,27.9,95.4)$                & $-2.1\pm 0.3$ & 0.3 \\
& Cand11b  & 23        & $(9.0,-0.1,2.4)$  & $(1.9,1.1,2.8)$                      & $(-152.7,80.2,258.2)$    & $(122.1,21.0,38.9)$               & $-2.0\pm 0.3$ & 0.5 \\
& Cand12   & 36        & $(9.6,-0.8,3.7)$  & $(2.0,2.4,4.2)$                      & $(-43.3,102.4,50.0)$     & $(172.8,21.2,197.8)$              & $-1.6\pm 0.2$ & 0.6 \\
& Cand13   & 36        & $(9.1,1.0,3.1)$   & $(2.5,2.0,4.1)$                      & $(-2.1,-13.2,202.2)$     & $(215.7,28.1,215.9)$              & $-1.4\pm 0.2$ & 0.4 \\
& Cand14a  & 24        & $(11.9,0.2,1.8)$  & $(1.8,1.7,3.6)$                      & $(-168.0,166.7,-25.1)$   & $(29.1,27.9,82.7)$                & $-1.4\pm 0.2$ & 1.2 \\
& Cand14b  & 12        & $(10.7,0.3,1.4)$  & $(1.8,2.1,3.5)$                      & $(193.6,202.9,-5.7)$     & $(14.3,13.5,51.8)$                & $-1.5\pm 0.1$ & 0.7 \\
& Cand15a  & 12        & $(10.5,1.4,4.0)$  & $(1.9,2.1,3.9)$                      & $(-297.4,220.0,-49.9)$   & $(29.6,23.5,79.3)$                & $-1.5\pm 0.1$ & 1.2 \\
& Cand15b  & 7         & $(10.3,-0.3,2.4)$ & $(1.8,2.3,5.9)$                      & $(291.3,207.3,48.3)$     & $(20.2,10.4,68.7)$                & $-1.4\pm 0.1$ & 0.5 \\
& Cand16a  & 12        & $(8.7,0.5,3.9)$   & $(1.6,1.5,3.9)$                      & $(315.2,109.2,-12.5)$    & $(30.9,4.6,67.2)$                 & $-1.4\pm 0.2$ & 0.7 \\
& Cand16b  & 5         & $(8.9,2.8,-1.3)$  & $(1.3,2.1,3.2)$                      & $(-360.7,147.5,81.7)$    & $(26.7,9.2,76.3)$                 & $-1.4\pm 0.1$ & 0.9 \\
& Cand17   & 10        & $(9.5,-0.4,2.0)$  & $(1.0,0.9,2.5)$                      & $(127.6,68.0,339.4)$     & $(157.4,8.0,54.8)$                & $-2.1\pm 0.2$ & 0.7 \\
\hline\hline
 \end{tabularx}
\caption{Basic parameters of the Shards. 
Those containing two populations in phase space are divided into two items ``a'' and ``b''. The Shards are
  organised into five categories: S1, S2, Retrograde, Prograde and
  Low-Energy (the latter group listed in Table~\ref{tab:restofshards}
  for brevity). The significance of the location of the Sun is given
  in $\sigma$ in the final column: a smaller number implies that
  the substructure spatial distribution overlaps the Sun.}
\label{tab:shards}
\end{table*}

 We fit each Shard with a Gaussian mixture model based on available data: 
 $\mathbf{q} = \{X,Y,Z,v_R,v_\phi,v_z,[{\rm Fe/H}]\}$, i.e.,
 $D=7$ in Eq.\eqref{eq:gaussian}. We include spatial information here
 to improve the measurement of the trajectory of the stream and to
 allow us to estimate the velocity at the solar position. While a
 Gaussian is a good description of a stream's cross section, it is
 not usually used to describe the shape along the stream. 
 However, given the small extent of the SDSS-\Gaia
 footprint relative to the size of the apocentric radii of the stellar
 orbits, this model should be fine. Ultimately the process of including the stellar spatial positions is only to be able to remove those Shards which clearly do not intersect the solar position.
 
 Many clusters of stars in action space correspond to separated
 clusters in velocity space coordinates. So we vary the number of
 possible populations, $n$ in Eq.(\ref{eq:gaussian}). In
 Fig.~\ref{fig:Vtriangle}, we show the stellar velocity data for four Shards fitted to these models. We choose the two highest
 significance streams (S1 and S2) and two examples of retrograde and
 prograde substructures (Rg5 and Cand14 respectively). 
 
 To quantify how distant each Shard is from Earth, we also display $P(\mathbf{x}_\odot)$, the
 significance of the solar position relative to the spatial
 distribution of the stars. We take the solar position in
 galactocentric rectangular coordinates to be $\mathbf{x}_\odot = (8.2,
 0, 0.014)$ kpc~\cite{Binney:1997,McMillan:2017,2019arXiv190405721A}
 and calculate the significance of this position in the Gaussian fit to
 the distribution of stellar locations.
 
 To evaluate the most appropriate number of populations, we select
 the option which yields the minimum Bayesian information criterion,
 although in most cases this choice is visible by eye. For three of the
 four examples shown, we find that there are two clear
 subpopulations. These represent the leading and trailing tails of
 stellar streams, which can stretch over large distances and wrap the
 Galaxy multiple times.  The power of action
 space searches for substructure is that they allow a single object to be identified, whereas a
 search in velocity space would have typically found two distinct
 clusters. The Shards that do not exhibit this bimodal feature instead
 are observable because they are close to the pericentric passage of
 the stream. In these cases, we are positioned close to the maximum
 stellar density along the stream. Figure~\ref{fig:XYZ} visualises this idea. We plot the individual stellar positions and
 velocities for three Shards and integrate the orbits of those stars forward and backward
 in time by 100 Myr using {\tt MWPotential2014} in the \texttt{galpy} package~\cite{Bovy:galpy}.
 
 In the first two examples, we see that stars are on
 orbits that are grouped together tightly and are at a position very
 close to pericentre. In the bottom row of
 Fig.~\ref{fig:XYZ}, however, Cand14 has at least two wraps coinciding with
 the solar position. The stars in this case
 are located at positions much larger than the mean pericentre of their
 orbits. This illustrates the interpretation discussed above, that the
 Shards consist of (i) streams with pericentres close to the solar
 position, and (ii) streams with multiple wrappings coinciding with the
 solar position. 

 A complete summary of the fit to each of the main substructures is
 given in Table~\ref{tab:shards}. There is another, longer list of the
 lower-significance (and typically lower-energy) Shards which we have
 included in Appendix~\ref{sec:lowEshards}. There are two streams in
 the Shards, labelled S1 and S2. The remaining substructures tagged as ``Rg''
 and ``Cand'' correspond to the categories Retrograde and Prograde. We
 now extend the naming convention of Ref.~\cite{Myeong:Preprint} to
 indicate those streams with multiple wraps entering the survey
 footprint. We add the suffix ``a'' and ``b'' in
 those cases. This subtlety is a key difference between this list and
 that of Ref.~\cite{Myeong:Preprint}. Accounting for multiple wraps
 prevents us overestimating the velocity dispersions of the substructures
 which could be anomalously large otherwise.

 In Table~\ref{tab:shards}, we also listed $P(\mathbf{x}_\odot)$, the
 significance of the solar position relative to the spatial
 distribution of the stars. In the next Section, we will devise a model for the Dark Shards in the
 velocity distribution. For this, we select only those candidates with
 significances $<2$ and populations $>4$. This means that only the
 most important Shards for DM searches are considered.  
Notably C2, which has a relatively high significance in action space, is located too far away from the Sun to be included in our model. 
 The Shards that remain are those listed in Tables~\ref{tab:shards}
 and~\ref{tab:restofshards}, a total of 59.

To streamline our discussion even further, we organise the Shards into five categories: S1, S2, Retrograde,
 Prograde and Low Energy. S1 and S2 are given their own status to
 reflect their very high significance as clusters in action space. The
 Retrograde Shards all have similar significances that are lower than
 S1 and S2, but since they share kinematic and chemical properties it
 is reasonable to group them together. This is also the case for the
 Prograde category. The Low Energy category consists of all of the
 remaining lower significance Shards listed in Table~\ref{tab:restofshards}.

 \section{Dark Shards in the halo model}\label{sec:darkshards}

The DM associated with the Shards---the Dark Shards---will
change the local velocity distribution of DM and therefore impact
the observable signals in direct detection experiments on Earth.  In
this section, utilising the properties of the Shards in
Table~\ref{tab:shards}, we construct a simple model for the Milky Way DM halo
to study the effects of the Dark Shards.

  Our model consists of three components: the roundish, isotropic halo
  $\fR (\vbf)$; the \Gaia Sausage $\fS(\vbf)$; and the Dark Shards
  $f_\xi(\vbf)$.  The velocity distribution in the {\it galactic frame}
  is the sum of the three components,
  \begin{equation}\label{eq:shmpp}
    f_{\rm{gal}}(\vbf) = \xi_{\rm{R}} \fR (\vbf) + \xi_{\rm{S}}  \fS(\vbf) +\sum_{i=1}^{n_{\xi}} \xi_i f^i_\xi(\vbf)\, ,
  \end{equation}
  where the sum extends over the $n_{\xi}$ Dark Shards and
  $\xi_{\rm{R}}$, $\xi_{\rm{S}}$ and $\xi_i$ give the fractional
  weighting to the local DM density in the Round Halo,
  Sausage and each Shard, respectively. By definition,
  $\xi_{\rm{R}}+\xi_{\rm{S}}+\xi_{\rm tot} =1$, where $\xi_{\rm tot} =
  \sum_{i=1}^{n_{\xi}} \xi_i$.  We will first discuss the functional
  forms for $\fR (\vbf)$, $\fS(\vbf)$ and $f^i_\xi(\vbf)$ before turning
  to discuss the weighting that we assign to each component.

  \subsubsection{The Round DM Halo}

  We model the velocity distribution of the Round DM Halo as a
  smooth, isotropic Gaussian distribution~\cite{Drukier:1986tm}:
  \begin{equation}\label{eq:shm}
  \begin{split}
  \fR(\vbf) = \frac{1}{(2\pi
    \sigma_v^2)^{3/2}N_\mathrm{R,esc}} \, &\exp \left( - \frac{|\vbf|^2}{2\sigma_v^2}\right) \\
    &\qquad \times \Theta (\vesc - |\vbf|)\,.
    \end{split}
   \end{equation}
  Here,~$\sigma_v$ is the velocity dispersion which is set, at the solar
  position, by the value of the local standard of rest~$v_0 = \sqrt{2}
  \sigma_v = 235$~km \!\!s$^{-1}$. We have adjusted the value of $v_0$
    from Ref.~\cite{Evans:2018bqy} by taking into account the more recent
    determination of the distance to the galactic centre, $R_0$~\cite{2019arXiv190405721A}. We truncate the velocity
  distribution at the galactic escape speed~$\vesc$, using the Heaviside
  function~$\Theta$.  The constant $N_\mathrm{R,esc}$ ensures the
  velocity distribution remains normalised to unity
  \begin{equation}\label{eq:norm}
  N_\mathrm{R,esc} = \erf \left( \frac{\vesc}{\sqrt{2}\sigma_v}\right) -
  \sqrt{\frac{2}{\pi}} \frac{\vesc}{\sigma_v} \exp \left(
  -\frac{\vesc^2}{2\sigma_v^2} \right)\, .
  \end{equation}
  A recent analysis that used a local sample of approximately $2300$
  high velocity counterrotating stars and a prior distribution inspired
  by numerical simulations obtained the value $\vesc =
  528_{-25}^{+24}$~km~s$^{-1}$~\cite{Deason19}, in agreement with, but
  improving upon, earlier determinations~\citep{Sm09,Williams:2017}.

\subsubsection{The Sausage}
We follow the \SHMpp~\cite{Evans:2018bqy} and model the Sausage component with a triaxial Gaussian velocity distribution,
\begin{align}\label{eq:sausage}
\fS(\vbf) =
\frac{1}{(2\pi)^{3/2}\sigma_r\sigma_\theta^2 N_\mathrm{S,esc}}
\, \exp \left( - \frac{v_r^2}{2\sigma_r^2}
-\frac{v_\theta^2}{2\sigma_\theta^2}- \frac{v_\phi^2}{2\sigma_\phi^2}
\right) \nonumber\\ \times \Theta (\vesc - |\vbf|)\, .
\end{align}
The velocity dispersions are related to the local standard of rest, $v_0 =235$~km \!s$^{-1}$, by~\citep{Ev97}
\begin{equation}\label{eq:disps}
  \sigma_r^2 = \frac{3v_0^2}{2(3-2\beta)},\qquad\sigma_\theta^2 =
  \sigma_\phi^2 = \frac{3 v_0^2(1-\beta)}{2(3-2\beta)}\, ,
\end{equation}
where $\beta$ is the anisotropy parameter. We
assume that $\beta$ for the DM in the Sausage is the
same as the value for the stars, $\beta =
0.9$~\cite{Be18,My18}. The escape speed normalisation modifier for this distribution is~\cite{Evans:2018bqy},
\begin{equation}
\begin{split}
N_\mathrm{S,esc} = \erf \left( \frac{\vesc}{\sqrt{2}\sigma_r}\right) &-
\sqrt{\frac{1-\beta}{\beta}}
\exp \left(-\frac{\vesc^2}{2\sigma_\theta^2} \right) \\
&\quad \times \erfi \left( \frac{\vesc}{\sqrt{2}\sigma_r} \cdot \sqrt{\frac{\beta}{1-\beta}}  \right).
\end{split}
\end{equation}
Since the \SHMpp is a model that contains only $\fR$ and $\fS$, we refer to the case when $\xi_{\rm tot} =
0$ as the \SHMpp.

  \subsubsection{The Dark Shards}

  In the context of this paper, $f^i_\xi(\vbf)$ is the most
  important component in Eq.\eqref{eq:shmpp} as it models the Dark
  Shards.  We write each subcomponent as
  \begin{equation}\label{eq:DarkShardsfv}
  \begin{split}
  f^i_\xi(\mathbf{v}) &= \frac{1}{(8 \pi^3 \det{\boldsymbol{\sigma}_\xi^{i\,2}})^{1/2} N_{_\xi,\rm{esc}}}  \\
  &\qquad \times \exp\left(-(\mathbf{v} - \mathbf{v}_\xi^i)^T \frac{(\boldsymbol{\sigma}_\xi^i)^{-2}}{2} (\mathbf{v} - \mathbf{v}_\xi^i) \right)\\
  & \qquad \qquad \times \Theta (\vesc - |\vbf|)\;.
  \end{split}
  \end{equation}
  Compared with the Round Halo and the \Gaia Sausage, the Dark Shards do
  not have zero mean velocity in the galactic frame of
  reference. Instead, the triaxial velocity distribution is offset by
  the velocity of the moving substructure, $\mathbf{v}_\xi$.
  Furthermore, we allow for a general velocity dispersion
  tensor~$\boldsymbol{\sigma}_\xi$.  We include the Heaviside
  function to truncate the velocity distribution at the escape speed. In
  practice, as the subcomponents are peaked away from~$\vesc$ and as
  their velocity dispersion is relatively small, the effect of the
  truncation is negligible.
  
  \begin{figure*}[!t]
   \includegraphics[width=0.99\textwidth]{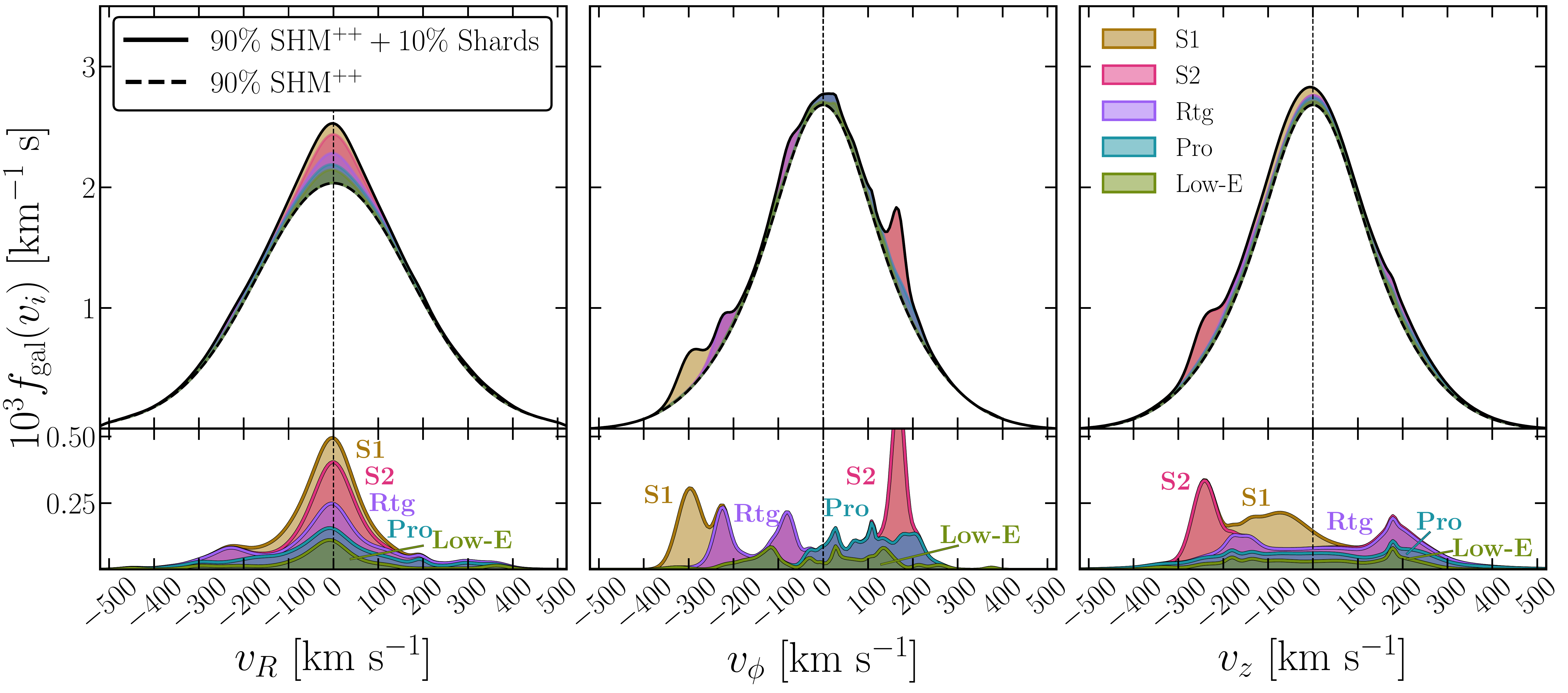}
  \caption{The distributions of the three components of the galactic
    rest frame velocities.  The solid black lines in the upper panels
    show the total distribution, including the isotropic Round Halo,
    \Gaia Sausage and Dark Shard components assuming the Dark Shards contribute 10\% to the local DM density ($\xi_{\rm{tot}}=10\%$).  
    The distribution without the Dark Shards is
    shown as the black dashed line.  The coloured regions show the
    contribution from each of the S1, S2, Retrograde, Prograde, and
    Low Energy categories: these distributions are stacked one upon the
    other in reverse order to aid the visibility of all the
    features. The lower panels show the stacked distributions from the
    Dark Shards alone.}
  \label{fig:Shards_fvgal}
  \end{figure*}

  We make the approximation that the velocity
  dispersions~$\boldsymbol{\sigma}_\xi$ that enter $f^i_\xi(\vbf)$ are
  the same as the stellar components listed in Table~\ref{tab:shards}.
  This assumption is probably not true in detail.  For instance, numerical
  simulations have shown that the DM stream associated with the
  Sagittarius stream are more extended and are misaligned from the
  stellar component~\cite{Purcell:2012sh,Purcell:2011nf}.  However, the
  progenitor of the Sagittarius stream is a dwarf irregular
  galaxy~\cite{NO10}, which had its stars initially distributed in a
  disk and its DM in a round halo, so this mismatch is an extreme case.
  The dwarf spheroidals, which are believed to have produced S1, S2, and
  the other Shards~\cite{Myeong:2017skt}, have DM and stellar
  populations that have similar spheroidal shapes before they were
  accreted.  Nonetheless, the velocity dispersion of the DM and stars in
  the progenitor is not exactly the same, and so mismatches between DM
  and stellar streams are still expected, albeit on a smaller scale than
  for the Sagittarius stream. Our approximation is reasonable
  for the dwarf spheroidal regime, though work to fully test its
  accuracy using a numerical simulations is desirable.

  \subsection{Weighting the halo components \label{sec:weight}}

  From Eq.\eqref{eq:shmpp}, we also need to specify $\xi_{\rm{R}}$,
  $\xi_{\rm{S}}$ and $\xi_i$, the fractional contribution that each
  component makes to the local DM density.  Although these cannot
  be determined from \Gaia data, there are some general statements that
  we can make about their relative values.  For example, the
  gravitational potential is nearly spherical~\cite{Wegg:2018voc}, so
  the DM associated with the Sausage and the Shards is subdominant. In
  Ref.~\cite{Evans:2018bqy} it was argued that the \Gaia Sausage can
  contribute around 20\% to the local DM density without
  exceeding the bound of 1\% on the ellipticity of the equipotentials of
  the Milky Way. A value $\sim 20\%$ is also consistent with the
  fraction obtained from the FIRE-2
  simulations~\cite{Necib:2018igl}, and is in accordance with the interpretation of the Sausage's formation determined using the Auriga simulations~\cite{2019MNRAS.484.4471F,Bozorgnia:2019mjk}.
    As the stellar Shards are not the dominant component of
  the main sequence stellar halo sample, it is unlikely that their
  progenitors will have brought more DM into the inner halo
  than, say, the Sausage. So $\xi_{\rm tot}$ will certainly be smaller
  than 20\%.

  To motivate us towards a lower limit to $\xi_{\rm tot}$ , we can refer
  to the library of N-body minor merger accretion
  events~\cite{Amorisco:2017}, as used in the interpretation of~S1
  and~S2 in Ref.~\cite{Myeong:2017skt}. Streams originating from
  $\sim10^{10}M_\odot$ and $\sim5\times10^9 M_\odot$ subhalos,
  believed to be the approximate total masses of the~S1 and~S2 progenitors,
  were found to contribute $\mathcal{O}(1\%)$ to the local DM density at
  the solar location. The simulations made simplifying approximations
  about the DM distribution in the in-falling satellite and the host
  galaxy, and the impact of the Milky Way disk was not included.
  Despite these limitations, they suggest that a reasonable range for
  the total contribution from all of the substructures in
  Tables~\ref{tab:shards} and~\ref{tab:restofshards} is $1\% \lesssim
  \xi_{\rm tot} \lesssim 10\%$. Following Ref.~\cite{Evans:2018bqy}, we
  will fix $\xi_{\rm{S}}=20\%$ and the round DM halo contributes the
  rest: $\xi_{\rm{R}}=80\%-\xi_{\rm tot}$.

  After this, we are still left with the task of
  specifying $\xi_i$, the contribution from each individual Dark Shard.
  As discussed in Sec.~\ref{sec:shards}, to reflect the relative
  importance of the clusters of stars in action space, it is prudent to
  group the Shards into five categories: S1, S2, Retrograde, Prograde,
  and Low Energy components.  We next apply an equal weighting to each
  category (e.g.,\ each category contributes 2\% when $\xi_{\rm tot} =
  10\%$).  Finally, since the individual Shards within each category
  have different star counts and positions relative to the Solar System,
  we weight the contribution of the Shards within each category by the
  number of stars multiplied by the height of the spatial distribution
  of stars at the solar position.  

While the weighting prescription just described is a benchmark
case, we will see it captures much of the resulting 
phenomenology of the different categories of substructure. 
If slight changes to the weighting lead to particularly
interesting experimental signals however, then these will also be discussed.

  \subsection{The galactic rest frame distribution}

  In Fig.~\ref{fig:Shards_fvgal} we present the galactic rest frame
  distributions of each galactocentric cylindrical polar velocity
  component.  The upper solid black line shows the velocity
  distributions obtained when $\xi_{\rm tot} = 10\%$ and the weighting
  prescription just described is used.  The lower dashed black line
  shows the \SHMpp distribution normalised to 90\%. This is to highlight
  the effect of the 10\% contribution from the Dark Shards which adds on
  top of the 90\% \SHMpp distribution: the five coloured regions between
  the dashed and solid black lines show the individual contribution from
  each of the S1, S2, Retrograde, Prograde, and Low Energy categories.
  The lower panels in Fig.~\ref{fig:Shards_fvgal} emphasise the range of
  velocities to which each category contributes.  The enhanced radial
  anisotropy of the \SHMpp due to the Sausage is the reason that the
  $v_R$ distribution (left panel) is wider and shorter than the other
  distributions.

The S1 stream and the Retrograde Shards at large negative $v_\phi$ are
interesting as they will become more prominent when boosted into the
Earth's frame.  Given that the Earth moves with $v_\phi \sim
250~\kms$, S1 and the Retrograde Shards blow into the Solar System
with a large $v_\phi \sim550~\kms$.  Some of the
phenomenology associated with the S1 stream has already been studied
in Refs.~\cite{OHare:2018trr,Knirck:2018knd,Buckley:2019skk}.

The S2 steam and the Prograde Shards are interesting because they have
a relatively large positive $v_\phi\sim200~\kms$ meaning they almost
corotate with the Earth. Corotating DM will have much lower speeds
when boosted to the Earth's frame.  While many methods of detecting DM
are more sensitive to the high speed tail, axion haloscopes are
notable because the speed distribution at low speeds can also give
rise to distinctive experimental signals. Also important is the large
negative $v_z$ of S2.  This behaviour can be seen in
Fig.~\ref{fig:XYZ} (middle right panel), which showed that S2 passes
through the Milky Way disk from above. This means that S2, unlike S1,
is highly misaligned from the expected orientation of the DM wind. It
will therefore also be interesting from the perspective of directional
and time dependent searches for DM.

  \section{Dark matter speed distribution and flux on Earth\label{sec:labboost}}

  When observed on Earth, incoming DM particles are boosted
  from the galactic frame in which we have been working until now, into
  the lab frame.  The lab-frame velocity distribution is
  \begin{equation}
  f_{\rm{lab}}(\mathbf{v})=f_{\rm{gal}}(\mathbf{v}+\mathbf{v}_\odot  +  \mathbf{v}_\oplus)\;
  \end{equation}
  where $\mathbf{v}_\odot = (11.1, v_0+12.24,7.25) \kms$ is the sum of
  the local standard of rest and the peculiar velocity of the
  Sun~\cite{Schoenrich:2009bx,McMillan:2017}, and $\mathbf{v}_\oplus$ is
  the Earth's velocity around the Sun (we will neglect the Earth's daily
  rotation).  We will include the time dependence of $\mathbf{v}_\oplus$
  when we discuss modulation signals in
  Sec.~\ref{sec:modulations}. Until that section, the time dependence of
  the Earth's motion will not impact our results, so we fix
  $\mathbf{v}_\oplus = (29.4, -0.11, 5.90) \kms$, the value on March~9,
  which gives a speed distribution equivalent to the time averaged
  result.

  \begin{figure}[!t]
   \includegraphics[width=0.49\textwidth,trim={0cm 0cm 0cm 0cm},clip]{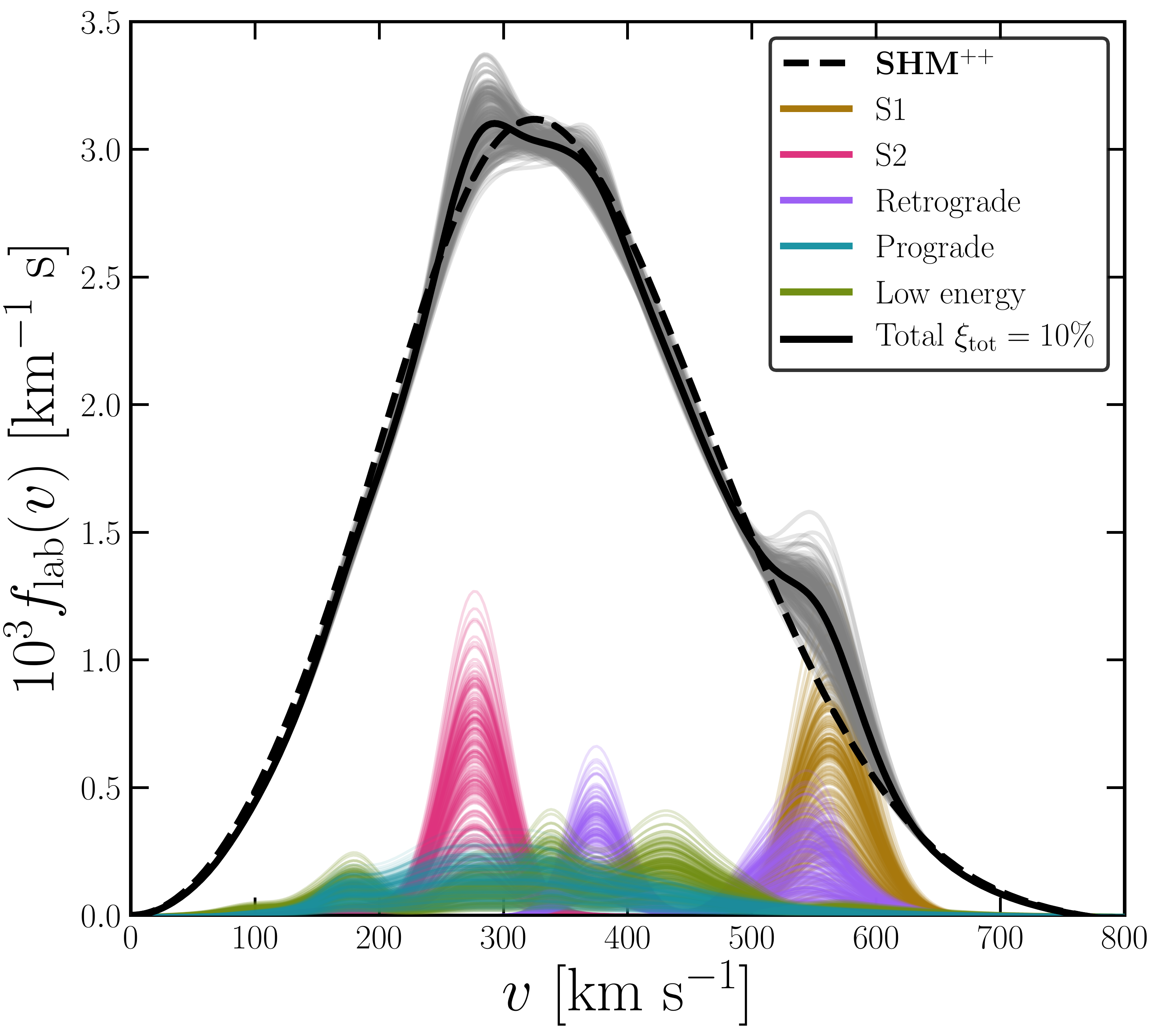}
  \caption{The lab-frame DM speed distributions.  The black
    solid line shows the distribution including the isotropic round
    halo, \Gaia Sausage and Dark Shard components assuming $\xi_{\rm
      tot}=10\%$ and the weighting prescription described in
    Sec.~\ref{sec:weight}.  For reference, the \SHMpp distribution is
    shown as a dashed black line.  The grey and coloured lines show the
    total speed distribution and individual contribution from each
    category (S1, S2, Retrograde, Prograde and Low Energy) obtained when
    the weightings of each category are selected randomly (such that
    $\xi_{\rm tot}=10\%$).  We see that the peak of the distribution
    (around $300\kms$) can be changed by S2 and the Prograde category while the high speed tail can be modified by
    S1 or the Retrograde Shards.  }
  \label{fig:Shards_fv_lab}
  \end{figure}

  \begin{figure*}[!t]
   \includegraphics[width=0.99\textwidth,trim={0cm 0.5cm 0cm 0cm},clip]{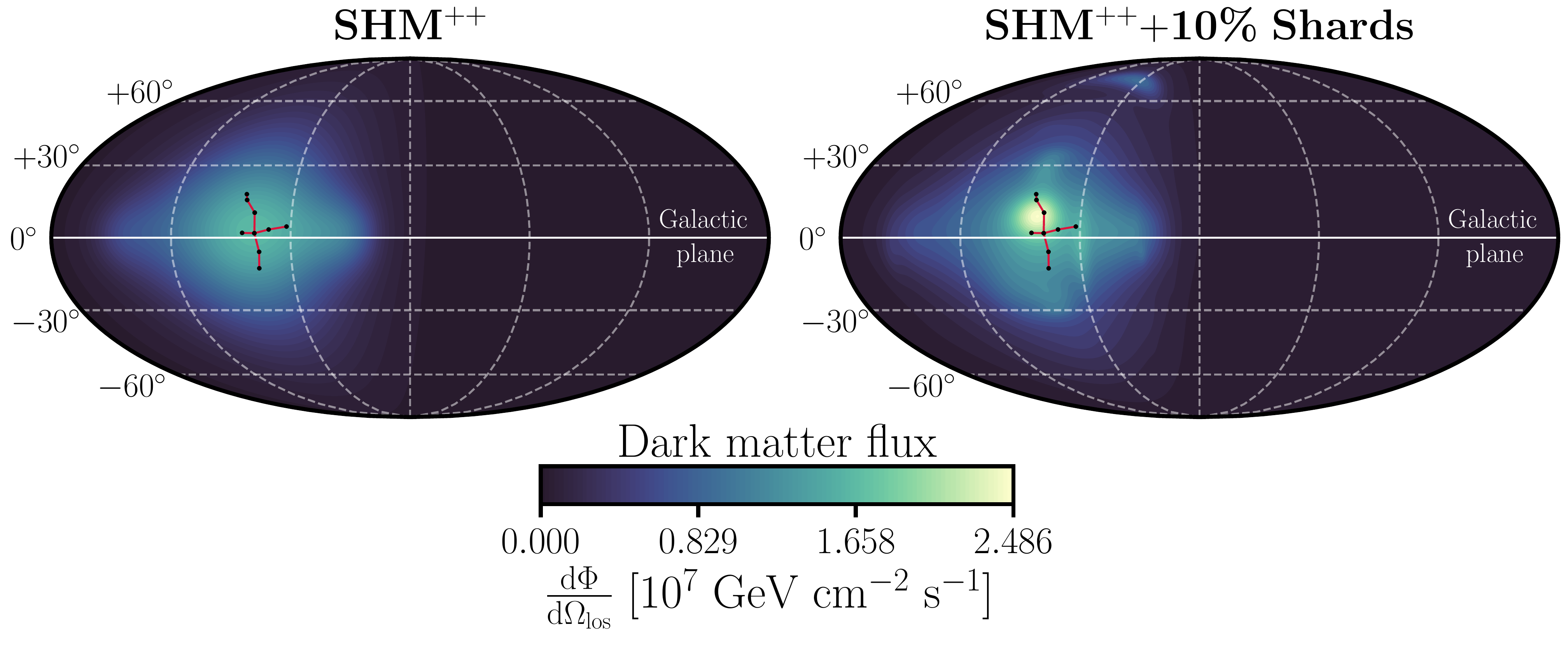}
  \caption{Angular dependence of the DM flux on Earth, in terms
    of galactic longitude and latitude. The galactic plane is shown by the white
    horizontal line and the Cygnus constellation in red.  The left panel
    shows the flux for the SHM$^{++}$, while the right panel shows the
    flux when the Dark Shards contribute $\xi_{\rm tot} = 10\%$.  In
    contrast to the SHM$^{++}$, which peaks in the direction of Cygnus
    and has a smooth distribution, the effect of the Dark Shards on the distribution is to
    enhance and slightly offset the peak towards Cygnus and to generate
    multiple subsidiary peaks.}
  \label{fig:Shards_DMFlux}
  \end{figure*}


  \subsection{Speed distribution}
  
Figure~\ref{fig:Shards_fv_lab} shows the DM {\it speed
  distribution} in the lab frame.  The black solid line shows the
speed distribution, $f_{\rm{lab}}(v)$, that is obtained when
$\xi_{\rm{tot}}=10\%$ and the components are individually weighted
according to the prescription described in Sec.~\ref{sec:weight}.  The
black dashed line shows the speed distribution from the \SHMpp model
(equivalent to setting $\xi_{\rm tot} = 0$).  The effect of~S1,~S2 and
the Retrograde structures are particularly apparent, as they enhance
the distribution around $280\kms$ and $550\kms$. 

To explore the possible effects of different weightings, we have also shown the distributions that are obtained when the
weights for the five substructure categories are drawn randomly from a
uniform distribution, while still adding up to
$\xi_{\rm{tot}}=10\%$. The effect on the overall speed distribution is
shown by the thin grey lines in Fig.~\ref{fig:Shards_fv_lab}, while the
contribution from the individual categories are shown by the coloured
lines at the bottom of the figure. Some care should be taken when
interpreting the envelope of the grey lines since, for example, if the contribution
from S2 is maximal ($\xi_{\rm{S2}}=10\%$) so that the peak at
$\sim280\kms$ takes the largest value, then there will be no peak at
$\sim550\kms$ as the contribution from the other categories is zero.
However, this exercise is useful as it highlights the general
features that can emerge and that are captured by our benchmark
weighting prescription; namely, the peak of the distribution (around
$300\kms$) can be changed depending on the weighting given to S2 and
the Retrograde categories, and the high speed tail can be modified
either by S1 or the Retrograde categories. The effects from the Prograde and Low Energy categories are
less pronounced in comparison to S1, S2 and the Retrograde Dark
Shards, but in general lead to a broad enhancement for speeds $v\lesssim 400 \kms$.

  \subsection{Directional dependence of the flux}\label{sec:DMfluxEarth}

In addition to the speed distribution, the directionality of
the DM signal is important for many experiments.  The
prediction of an anisotropic flux of DM is a generic result that
follows from our motion through the (essentially) non-rotating DM
halo.  For the simplest models in which the velocity ellipsoid (in the
galactic frame) is centred at $\mathbf{v} = \mathbf{0}$ and there is
no overall rotation, the DM flux on Earth peaks in the
direction~$-\vlab$.  This holds for both the isotropic part of the halo and
the Sausage, so the flux is largest in the
direction~$-\vlab$~\cite{Evans:2018bqy}, towards the constellation Cygnus. For substructures, this need not be the case. Since they may be characterised by nonzero average velocity in the rest
frame of the Galaxy [cf.\ Eq.\eqref{eq:DarkShardsfv}], their
lab-frame distribution will not necessarily align with Cygnus.

  Figure~\ref{fig:Shards_DMFlux} shows the angular distribution of the
  DM flux as seen on Earth.  This is calculated by integrating
  along the line of sight~$\hat{\mathbf{x}}_{\rm los}$,
  \begin{equation}
  \frac{\textrm{d}\Phi_{\rm DM}}{\textrm{d}\Omega_{\rm los}} = \rho_0 \int_0^{\infty}\, v^3 \, f_{\rm lab}(-v\, \hat{\mathbf{x}}_{\rm los})\, \mathrm{d} v \, ,
  \end{equation}
  where $\rho_0=0.55~\mathrm{\GeV}/\mathrm{cm}^3$~\cite{Evans:2018bqy} is
  the local DM density. For this result and several others, we display
  angular distributions in terms of galactic longitude and latitude, $(l,b)$,
  which are mapped with a Mollweide projection.
  
  The left panel shows the flux for the \SHMpp with no substructure (i.e., $\xi_{\rm tot} = 0$).
The distribution is smooth and peaks towards Cygnus, but the contours of equal flux
are not quite symmetric owing to the anisotropy brought by the Sausage. 
The right panel shows the flux when the Dark
Shards contribute~$\xi_{\rm tot} = 10\%$ of $\rho_0$. Compared to the left
panel, we see that the peak of the flux is now offset slightly from
the direction of Cygnus, the maximum flux is much higher, and
there are multiple peaks rather than one, with the most
noticeable one appearing towards high galactic latitudes, $b \approx +70^\circ$.

\begin{figure}[!t]
   \includegraphics[width=0.49\textwidth,trim={0cm 0cm 0cm 0cm},clip]{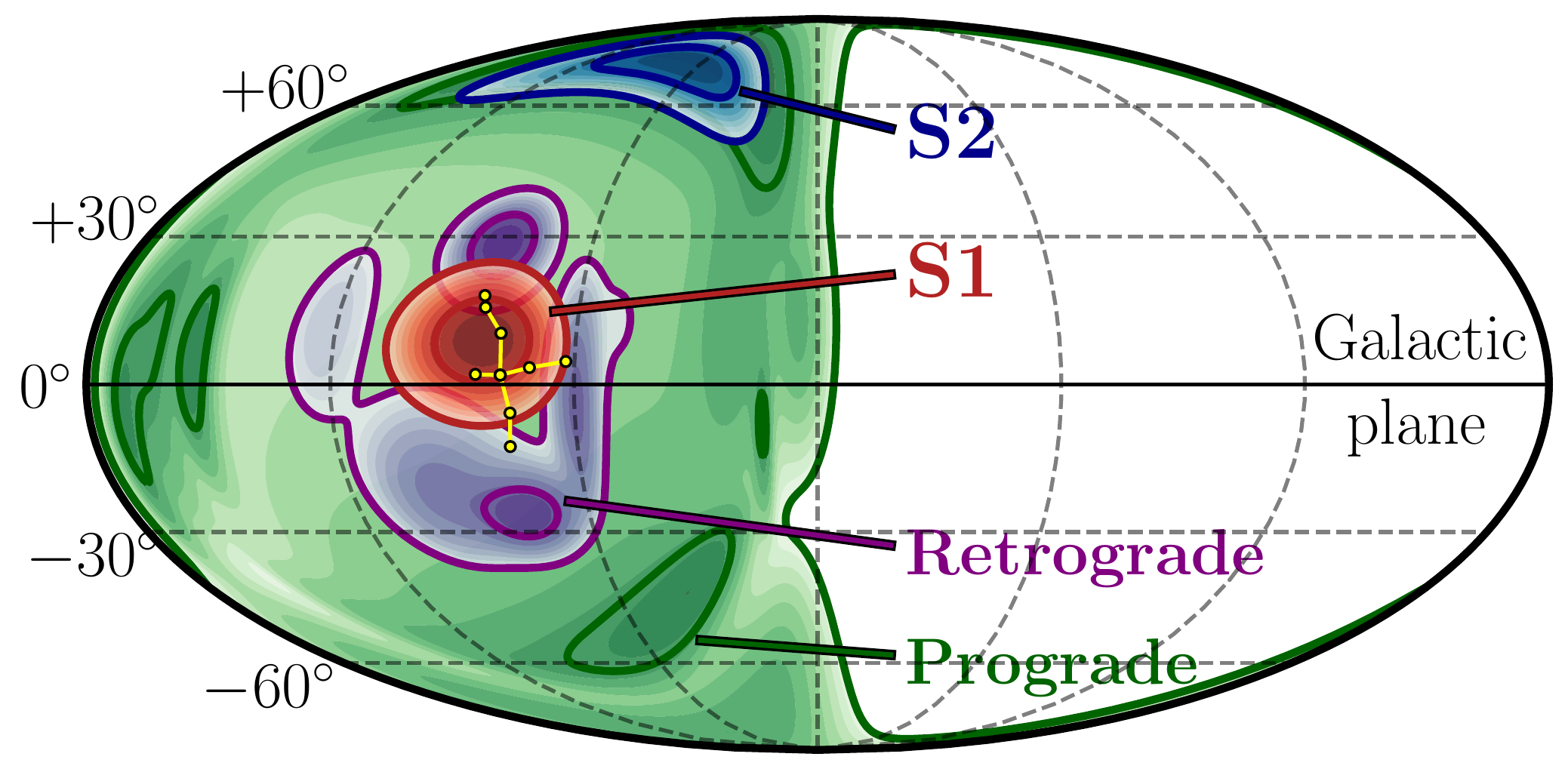}
  \caption{Angular distribution of the Dark Shards (with the exception of the
    Low Energy category).  For each component we delineate 
    contours enclosing 68\% and 95\% of the flux distribution in angle. We see that S1 and the Retrograde
    Shards are focused around the Cygnus constellation, while the Prograde and S2 Shards
    are spread over much wider angles.}
  \label{fig:ShardsInTheSky}
  \end{figure}

  To understand the origin of these differences, it is helpful to view
  the angular distributions of the individual categories.  These
  are shown in Fig.~\ref{fig:ShardsInTheSky}, where we have drawn the
  68\% and 95\% contours around each category's line-of-sight flux distribution.  The S1 stream almost aligns with Cygnus
  because of its relatively small $v_R$ and $v_z$
  (cf.~Table~\ref{tab:shards}). The
  Retrograde Shards are also focused more strongly around Cygnus as the
  boost into the Earth-frame increases the size of their $v_{\phi}$ components relative to
  $v_R$ and $v_z$.  This is reversed in the case of the Prograde category, whose flux becomes spread
  over much wider angles relative to Cygnus.  Finally, the prominent high
  latitude feature from Fig.~\ref{fig:Shards_DMFlux} can clearly now be
  identified with S2. This should not be surprising given the large negative
  $v_z$ component that we observed in Fig.~\ref{fig:Shards_fvgal},
  indicating the presence of a component incoming from high latitudes.  For visibility, we
  have not shown the Low Energy Shards which, similarly to the Prograde
  structures, are spread over much wider range of angles.

  \section{Axion searches}\label{sec:axions}

  The changes we have made to the lab-frame speed distribution are
  relevant for all experiments searching for DM, but
  particularly axion searches. Since axions are so light, to make up the nonrelativistic galactic DM the axion field would need to have enormous, macroscopic occupation numbers per quantum state. As such the natural description of axionic DM is of an oscillating classical field, which for small enough time and length-scales can be written,
  \begin{equation}\label{eq:axionfield}
  a(t) = \frac{\sqrt{2\rho_a}}{m_a} \cos{\left[ m_a\left(1+ \frac{v^2}{2}\right) t - m_a \mathbf{v}\cdot\mathbf{x} + \phi\right]}\;,
  \end{equation}
  where $m_a$ is the axion mass, $\mathbf{x}$ is position, and $\phi$ is
  an arbitrary phase. The amplitudes of the oscillations of the axion field are expressed in terms of the axion density $\rho_a$, which is
  stochastically varying under an exponential distribution with mean
  $\rho_0$ (see e.g.,\ Refs.~\cite{Centers:2019dyn,Derevianko:2016vpm,
    Knirck:2018knd} for further details).

  The axion field is coherent according to Eq.(\ref{eq:axionfield}) for
  timescales of the order $\tau_a \lesssim (m_a v^2)^{-1}$, where $v$ is a
  ``typical'' DM speed. The axion
  oscillations will decohere over timescales longer than $\tau_a$ and
  the subsequent temporal variation in frequency and phase will
  effectively map out the DM speed distribution when observed
  for long enough. Haloscope designs generically
  involve the monitoring of some electromagnetic response to $a(t)$. The
  spectral density of photons (usually the Fourier transform of
  time series data) measured over a time $t\gg \tau_a$ is
  proportional to $f_{\rm{lab}}(v)$ up to stochastic variations. Hence the signal lineshape in axion
  haloscopes is very sensitive to the DM halo model.

  The central problem in detecting the axion however is that we do not
  know the frequency, $\omega\simeq m_a(1+v^2/2)$ at which the
  electromagnetic response to the axion field should be monitored. To
  search for this frequency, haloscopes either enforce a resonance or
  constructive interference condition for a signal oscillating at
  $\sim m_a$ (as in e.g.,~ADMX~\cite{Asztalos:2009yp,Du:2018uak},
  MADMAX~\cite{TheMADMAXWorkingGroup:2016hpc,Millar:2016cjp},
  HAYSTAC~\cite{Brubaker:2016ktl,Rapidis:2017ytq,Brubaker:2017rna,Zhong:2017fot},
  CULTASK~\cite{Chung:2016ysi,Lee:2017mff,Chung:2017ibl},
  ORGAN~\cite{McAllister:2017lkb,McAllister:2017ern},
  KLASH~\cite{Alesini:2017ifp} and RADES~\cite{Melcon:2018dba}), or are
  sensitive to a wide bandwidth of frequencies simultaneously
  (e.g.,\ ABRACADABRA~\cite{Kahn:2016aff,Ouellet:2018beu,Ouellet:2019tlz},
  BEAST~\cite{McAllister:2018ndu} and
  DM-Radio~\cite{Silva-Feaver:2016qhh}). See Ref.~\cite{Irastorza:2018dyq} for a recent review.
  
     \begin{figure}[t]
  \centering
  \includegraphics[width=0.49\textwidth]{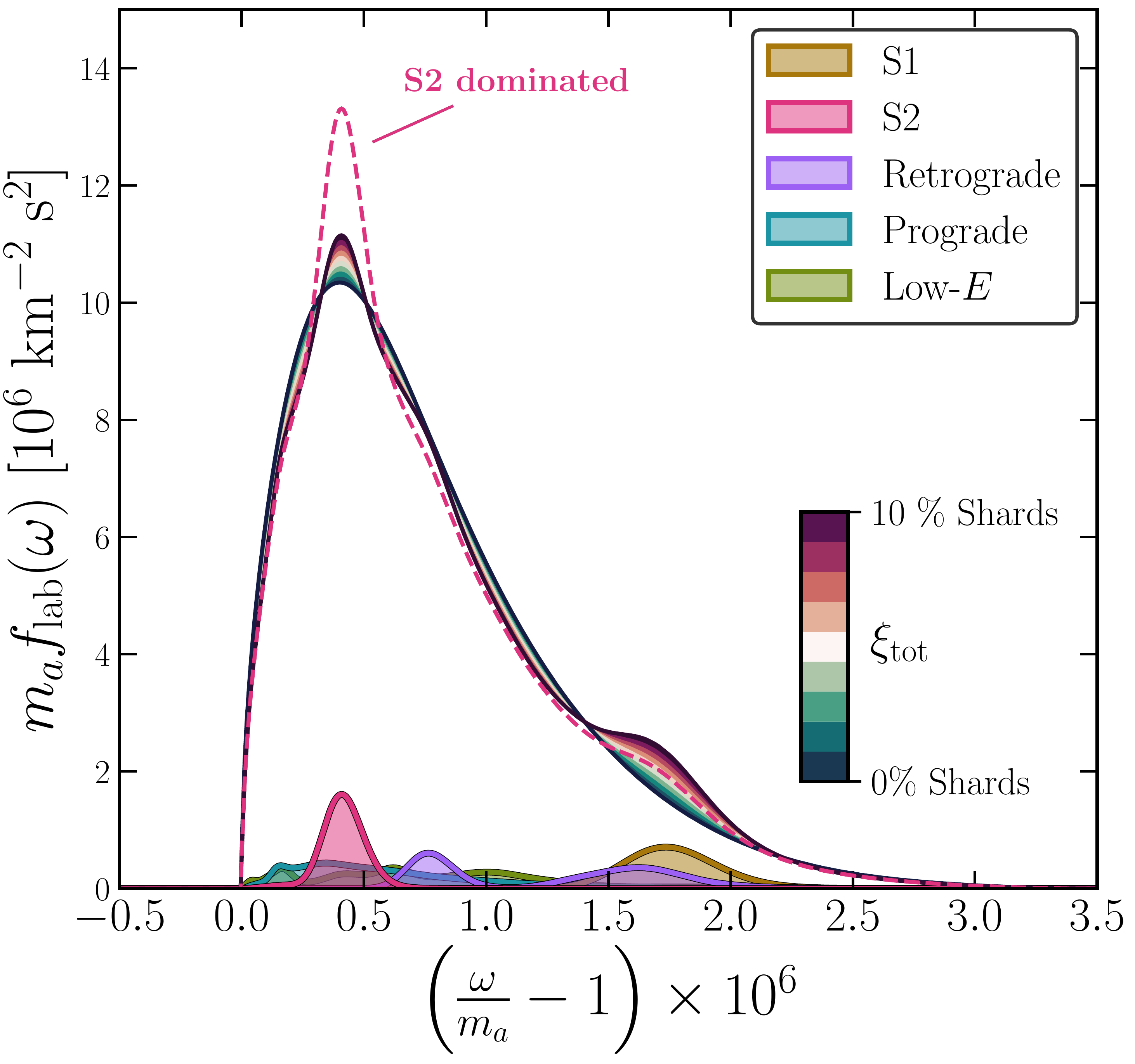}
  \caption{The time-averaged spectral lineshape observable in an axion
    haloscope as a function of frequency $\omega$ relative to the axion
    mass $m_a$.  The lineshape from the \SHMpp + Dark Shards equal
    weighting halo model is shown by the coloured region, from dark
    green to dark red corresponding to the range of values $0\%\leq
    \xi_{\rm tot} \leq 10\%$. We also show the individual weighted
    contribution from each Shard category when $\xi_{\rm tot}=10\%$. S1
    and the Retrograde Shards contribute to a bump at
    higher frequencies but S2 sharpens the peak of the
    distribution. To emphasise this, the pink dashed line shows the lineshape under an alternative weighting
    scheme in which half of $\xi_{\rm tot}=10\%$ is assigned to S2.
    This distribution is notably sharper and would enhance the
    sensitivity of axion haloscopes to the coupling
    $g_{a\gamma}$.}\label{fig:axionspectrum}
   \end{figure}

  For most of the haloscope techniques discussed, the spectral density of axion-induced photons is proportional to the speed distribution,
  up to a change of variables between frequency and speed (see
  e.g.,\ Refs.~\cite{OHare:2017yze,Foster:2017hbq})
  \begin{equation}\label{eq:axionpower}
  \frac{\textrm{d}P}{\textrm{d}\omega} = \pi \mathcal{H}(\omega) \,g^2_{a\gamma} \,\rho_0 \, f_{\rm{lab}}(\omega) \, ,
  \end{equation}
  where $g_{a\gamma}$ is the axion-photon coupling on which the experiment
  will set a limit.  We use $\mathcal{H}(\omega)$ to note that a haloscope will often have an underlying frequency dependence, but over the small width of the axion signal this is usually going to be flat. The relevant object for us is,
  \begin{equation}
  f_{\rm{lab}}(\omega) =\frac{\textrm{d}v}{\textrm{d}\omega}\, f_{\rm{lab}}(v)\, .
  \end{equation}
  We show $f_{\rm{lab}}(\omega)$ as a function of frequency in
  Fig.~\ref{fig:axionspectrum}.  The coloured shading from dark green to
  dark red indicates the changes to the lineshape as
  $\xi_{\mathrm{tot}}$ is increased from 0\% (the \SHMpp) to 10\%.  The
  contribution from the five Dark Shard categories when
  $\xi_{\mathrm{tot}}=10\%$ and the equal weighting scheme
  (cf.\ Sec.~\ref{sec:weight}) is shown at the bottom of the figure.  We
  see similar effects to those shown in Fig.~\ref{fig:Shards_fv_lab}, but with an
  important difference: since $\omega \propto v^2$, the lineshape is
  more concentrated around low values than the speed distribution shown earlier.  In
  fact, the peak itself is notably sharper relative to the \SHMpp due to S2.  
  However overall, the lineshape ends up being slightly wider due to the
  presence of S1 and the Retrograde Shards at higher frequencies. The
  effect from S1 was considered in detail in Ref.~\cite{OHare:2018trr},
  including also the dependence on the S1 velocity dispersion.

  In both resonant and broadband haloscopes, the sensitivity to
  $g_{a\gamma}$ is dependent upon how prominently the signal can show up
  over the experiment's noise floor. In a generic statistical
  methodology, this means that the sensitivity of an axion experiment
  scales as $g_{a\gamma} \sim (\int
  f(\omega)^2\,\mathrm{d}\omega)^{-1/4}$. Signals that are sharper in
  frequency are more prominent over white noise and hence easier to
  detect. Since the effect from S2 appears at the peak of the lineshape,
  if the local DM density had a larger contribution from S2, we
  would expect an even sharper lineshape.  To demonstrate this
  explicitly, the pink dashed line in Fig.~\ref{fig:axionspectrum} shows
  the lineshape when half of $\xi_{\rm tot} = 10\%$ is weighted towards
  S2 (as opposed to a fifth under the equal weighting scheme). This
  distribution is notably sharper and further increases the
  sensitivity of axion haloscopes to the coupling $g_{a\gamma}$. 
  
We calculate a 2\%--5\% enhancement in the experimental sensitivity relative to the \SHMpp when the fraction of $\rho_0$ attributed to S2 is between 5\%--10\%. 
The precise enhancement depends sensitively on S2's velocity dispersion, which --- since there is a contamination from outlying stars---may be overestimated. 
To obtain more robust results from axion haloscopes, it is therefore important that the properties of S2 are characterised precisely. 
For instance, if the S2 velocity dispersion is reduced by 20\% then we would expect a 10\%--20\% enhancement in axion sensitivity. 
Although this is a modest enhancement, as the axion mass scan rate to reach a given coupling scales as $g^{-4}_{a\gamma}$, it implies a reduction in the running times for reaching the sought-after DFSZ axion models~\cite{Dine:1981rt,Zhitnitsky:1980tq}.
In fact the S2 enhancement is close to the enhancement brought about by the ``N-body inspired'' line shape often used in the presentation of results from ADMX (see e.g.\ Ref.~\cite{Braine:2019fqb}). This enhancement is modulo the increase in the local density between that model and the SHM.
  
Once the axion has been detected, it becomes more straightforward to
measure the properties of streams. Since there will no longer be any
need to scan the axion mass parameter, the experiment can reach a
high statistics regime very rapidly and the signal will continue to
amplify above the noise. For example, for ADMX to achieve
 DFSZ-sensitivity with signal-to-noise $S/N=5$ over a
window of $20~\upmu$eV in 5 years, around $\sim10^6$ mass bins would
be needed, corresponding to a typical time of $T\sim100$~s on each
bin. To subsequently measure another peak in the spectrum which has an
amplitude that was some fraction of the main signal, the noise would need
to be reduced by that same ratio. Keeping the frequency resolution
the same, the amplitude of white noise will scale with
$1/\sqrt{t}$. The S2 peak is around a fraction $0.23(\xi_{\rm
  S2}/5\%)$ of the height of the main peak. Therefore we can estimate
that around $(100\,{\rm s}/0.23)^2 = 52$ hours would be required to
measure S2 to the same $S/N$ as required to initially detect the DFSZ
axion. From this measurement, the density of S2
relative to the rest of the halo can be determined from the
height of the peak, as well
as the lab frame speed and dispersion along one dimension from its
central frequency and width.
Furthermore, from the phase and amplitude of the
annual modulation of S2, the rest of the 3-dimensional dispersion and
velocity could be measurable on the timescale of a year. This procedure
is discussed in much greater detail in Ref.~\cite{Knirck:2018knd}.
A postdiscovery haloscope can be considered to be a very high-statistics 
instrument due to the alleviation of the need to scan over the axion mass.

 \subsection{Dependence on the dark matter direction}\label{sec:axiondirection}

 Some classes of axion experiments are also sensitive to the directionality of
  the DM flux.  The CASPEr experiments for example~\cite{Budker:2013hfa}
  utilise spin-precession to detect the nuclear coupling of
  ultralight axions.  The generic Hamiltonian that CASPEr is sensitive
  to has the form $\mathcal{H} \sim g\, \mathbf{I}_N \cdot \mathbf{D}$,
  where $\mathbf{I}_N$ represents the polarised nuclear spins and
  $\mathbf{D}$ is an effective field.  The
  CASPEr-wind~\cite{Garcon:2019inh} experiment assumes that the
  effective field is given by the spatial gradient of the axion field,
  \begin{equation}
  \mathbf{D}_a(t) \simeq -\sqrt{2\rho(t)} \sin(m_a t + \phi) \, \mathbf{v}(t) \, .
  \end{equation}
  In this case, the Hamiltonian is proportional to the scalar product of
  the axion velocity and the polarised nuclear spin so the experiments
  are most sensitive when these two vectors are aligned. For higher mass axions (and electron and photon couplings), the ferromagnetic haloscope QUAX~\cite{Crescini:2018qrz,Crescini:2016lwj,Barbieri:2016vwg,Barbieri:2016vwg} also measures an effective field dependent on the axionic gradient. 
  
  One typically assumes a smooth DM flux that on average points in
  the direction of Cygnus $\hat{\mathbf{x}}_{\rm Cyg}$.  Yet we saw in
  Fig.~\ref{fig:ShardsInTheSky} that one of the effects of the Dark
  Shards was to displace slightly the peak of the DM flux from
  the direction of Cygnus and to introduce a prominent high latitude
  component due to S2.  This means that the gradient of the axion field as it varies over the coherence length and time will therefore be more likely to point at large angles away from Cygnus than under the assumption of the SHM. This will modify the daily modulation~\cite{Knirck:2018knd}, and may potentially affect experimental sensitivities for CASPEr and QUAX. Similar arguments may also apply to experimental methods involving atomic clocks and comagnetometers that are searching for a wider class of ultralight particles~\cite{Afach:2018eze,Alonso:2018dxy,Wolf:2018xlz}. We leave a more detailed investigation of this subject to future work.

  \section{Nuclear recoil signals}\label{sec:recoils}

  \begin{figure}[t]
  \includegraphics[width=0.49\textwidth]{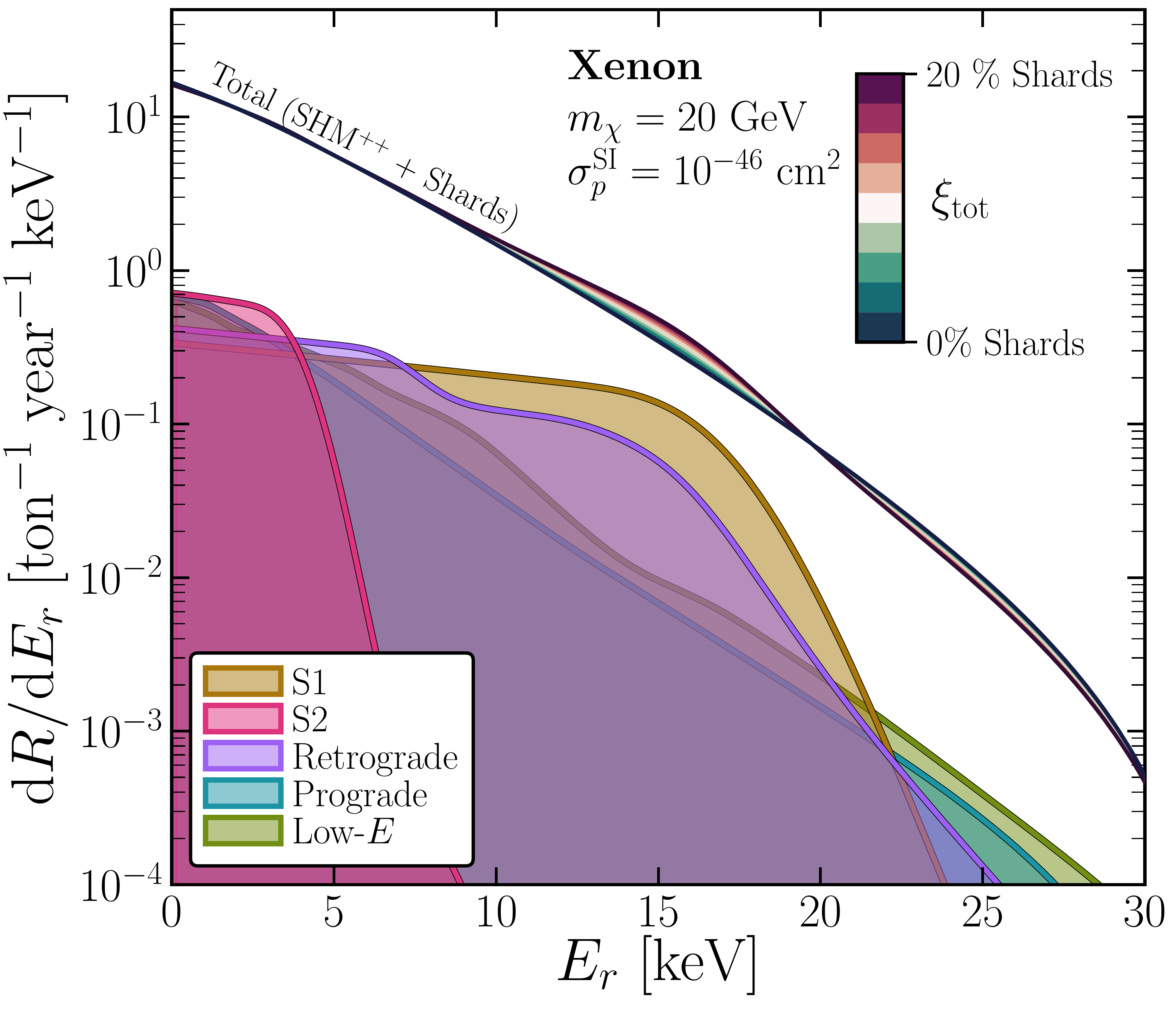}
  \caption{Time-averaged differential event rate as a function of energy
    for a~20~\GeV WIMP-like particle scattering through a
    spin independent interaction with xenon.  The coloured shading from
    green to red indicates how the event rate changes as the fraction
    $\xi_{\rm tot}$ is increased from 0\% to 20\%.  The coloured lines
    at the bottom of the figure show the individual contribution from
    each category: S1, S2, Retrograde, Prograde and Low Energy.  The
    dominant change in the spectrum comes from S1 around 15 \keV, although the effect is
    always small, even when we increase the maximum limit of $\xi_{\rm tot}$ to 20\% to make
    the effect more noticeable.}\label{fig:eventrates}
  \end{figure}

  Many DM experiments currently operating search for signals
  from WIMP-like particles from the Milky Way halo scattering with
  nuclei through a wide variety of
  interactions~\cite{Drukier:1983gj,Goodman:1984dc,Undagoitia:2015gya}.  For two-to-two
  scattering processes (which could be elastic or
  inelastic~\cite{TuckerSmith:2001hy,Baudis:2013bba,McCabe:2015eia,Arcadi:2019hrw}),
  the time-averaged rate~$R$ of nuclear scatters as a function of the
  energy of the recoiling nucleus~$E_r$ is found by integrating the flux
  of DM,
  \begin{equation}\label{eq:eventrate}
   \frac{\textrm{d}R}{\textrm{d}E_r} = \frac{1}{m_N} \frac{\rho_0}{m_\chi} \int_{v>\vmin} \!\!\!\!  v \, f_{\rm{lab}}(\mathbf{v}) \,\frac{\mathrm{d} \sigma_T (v,E_r)}{\mathrm{d} E_r}  \, \textrm{d}^3v \, .
  \end{equation}
  Here, $m_N$ is the target nucleus mass, $m_{\chi}$ is the WIMP-like
  particle mass, $v_{\rm{min}}$ is the minimum DM speed that can induce
  a recoil of energy~$E_r$, and $\sigma_T$ is the DM--nucleus scattering
  cross section, which can depend on $v$ and $E_r$.

  It is commonplace to extract all factors of $v$ from the cross section
  and absorb them into the halo integral, a single object that
  encapsulates all of the velocity distribution dependence.  For the
  canonical leading order spin independent (SI) and spin dependent (SD)
  DM--nucleus interactions, the differential
  cross section is inversely proportional to the square of the DM speed,
  $\mathrm{d}\sigma_T/\mathrm{d}E_r\propto v^{-2}$. In this case, the
  relevant halo integral is the mean inverse speed above $v_{\rm min}$,
   \begin{equation}\label{eq:gvmina}
  g(v_{\rm min}) = \int_{v>\vmin} \frac{f_{\rm{lab}}(\textbf{v})}{v} \,
  \textrm{d}^3 v \, .
   \end{equation}

  To show the impact of the Dark Shards on observable signals, we will
  consider the most familiar one: spin independent elastic scattering with equal
  couplings to protons and neutrons. In this case,
  \begin{equation}
  \frac{\mathrm{d} \sigma_T (v,E_r)}{\mathrm{d} E_r} = \frac{m_N A^2 \sigma_p^{\rm{SI}}}{2 \mu_p^2\,v^2} F^2(E_r)\;.
  \end{equation}
  Here, $A$ is the atomic number, $\mu_p$ is the DM-proton reduced mass,
  $F(E_r)$ is the nuclear form factor and $\sigma_p^{\rm{SI}}$ is the SI
  DM-proton scattering cross section.  We display the differential event
  rate in Fig.~\ref{fig:eventrates} for an experiment with a xenon
  target (applicable for LZ~\citep{Akerib:2018lyp},
  PandaX~\cite{Tan:2016zwf}, XENON~\citep{Aprile:2018dbl} and XMASS~\cite{Abe:2013tc,Suzuki:2018xek}).  The coloured shading from dark green to dark red
  indicates the changes to the differential rate as $\xi_{\mathrm{tot}}$
  is increased from 0\% (the \SHMpp) to 20\%.  The coloured regions at
  the bottom of Fig.~\ref{fig:eventrates} show the contribution from the
  individual Dark Shards.  The S1 stream and most of the Retrograde
  Shards have the largest impact on the differential rate. S2 on the
  other hand, while prominent in other signals, is much
  less important here.

  We find that the impact of the Dark Shards on
  $\mathrm{d}R/\mathrm{d}E_r$ is small, though this is not wholly
  unexpected. Xenon detectors, and nuclear recoil experiments in
  general, are poor at distinguishing features in the velocity
  distribution. Since only the recoil energy $E_r$ is measured while
  information about the recoil direction is lost, DM velocities cannot
  be reconstructed. Mathematically, this is encoded in the fact that the recoil rate is
  proportional to an integral over speeds, $g(v_{\rm min})$, rather than
  the speed distribution explicitly as in the case of the axion-induced
  photon spectral density (cf.~Eqs.~\eqref{eq:axionpower}
  and~\eqref{eq:eventrate}). 

  We displayed $\mathrm{d}R/\mathrm{d}E_r$ for a 20 \GeV WIMP-like particle
  here. For masses of this size scattering with xenon, the spectrum
  should in fact be the most sensitive to changes to $f(v)$
  (see Ref.~\cite{OHare:2018trr} for a quantitative discussion). However despite
  this choice, and even increasing the maximum value of
  $\xi_{\mathrm{tot}}$ to 20\% to make the changes more noticeable, they
  remain small.

  As most of the changes in the recoil spectrum are brought about
  by S1, the phenomenology of the Dark Shards as a departure from the
  SHM will be very similar to the case studied in
  Ref.~\cite{OHare:2018trr}. Following what was concluded then, we
  concur that the impact of the Dark Shards will be minor for nuclear
  recoil based experiments unless S1 and the Retrograde shards are particularly prominent in the DM of the {\it ex situ} halo.

  \subsection{Modulation signals}\label{sec:modulations}
  \begin{figure}[t]
  \includegraphics[width=0.49\textwidth]{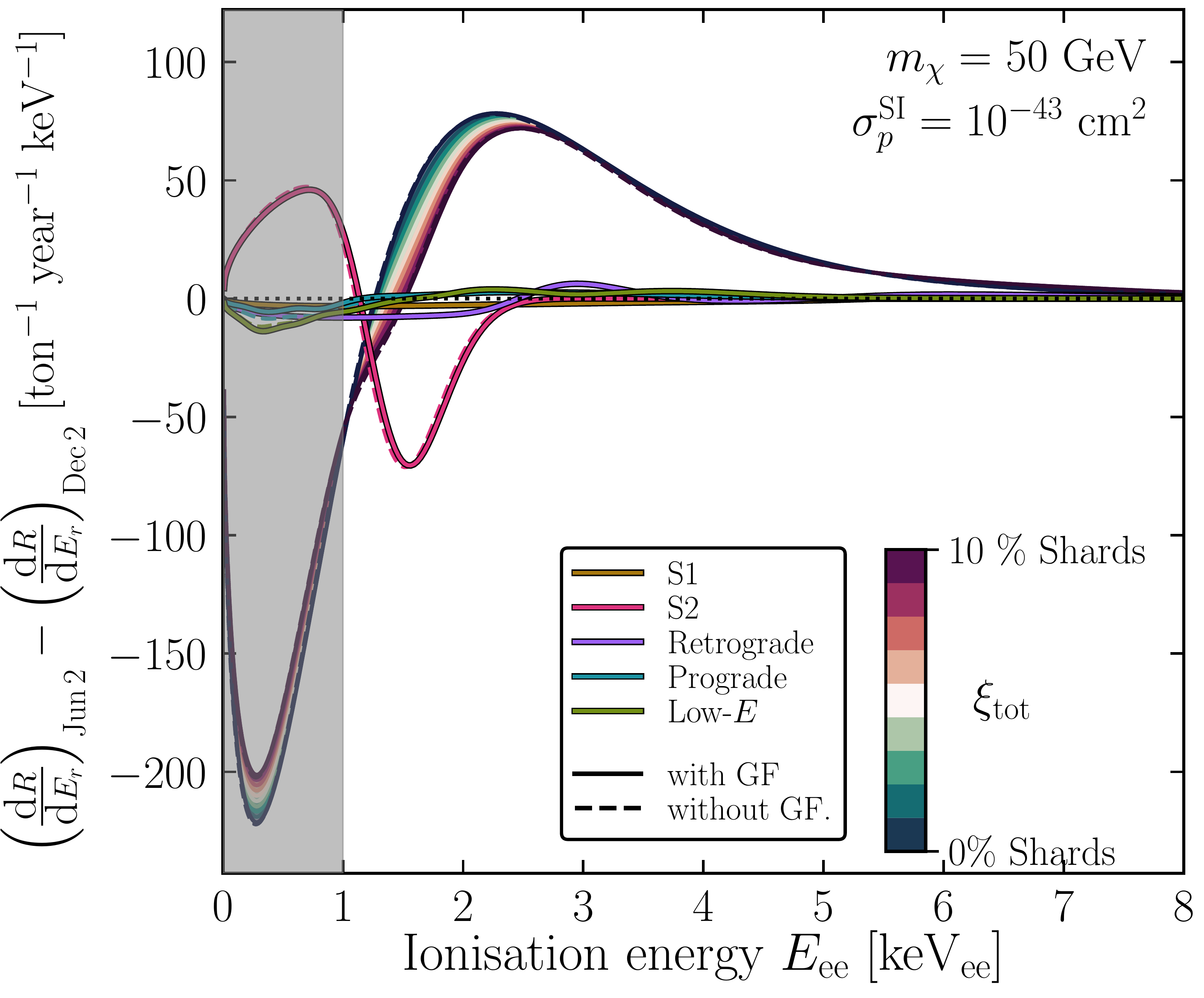}
  \caption{Modulation amplitude (defined as the difference in the rate
    between June 2 and December 2) versus ionisation energy in a NaI
    scintillator for a 50~\GeV WIMP-like particle and a spin independent
    interaction. The shaded band indicates the change in the modulation
    amplitude as the value of $\xi_{\rm tot}$ increases from 0 to 10\%
    (green to red). The modulation amplitude decreases as $\xi_{\rm
      tot}$ increases.  The coloured lines show the modulation amplitude
    from the individual Dark Shards categories assuming (for visibility)
    that each contributes 20\% of the local DM density. S2 is the most
    distinctive as its phase is reversed over most of this range of energies and its modulation amplitude is
    larger compared to the other Dark Shards.  The dashed lines show the
    modulation amplitude when gravitational focusing is not included.
    They lie very close to the solid lines
    indicating that the effect on {\it this}
    measure of the modulation amplitude is small.
  }\label{fig:Shards_annualmod_gravfocus}
  \end{figure}

  \begin{figure*}[t]
  \includegraphics[width=0.99\textwidth]{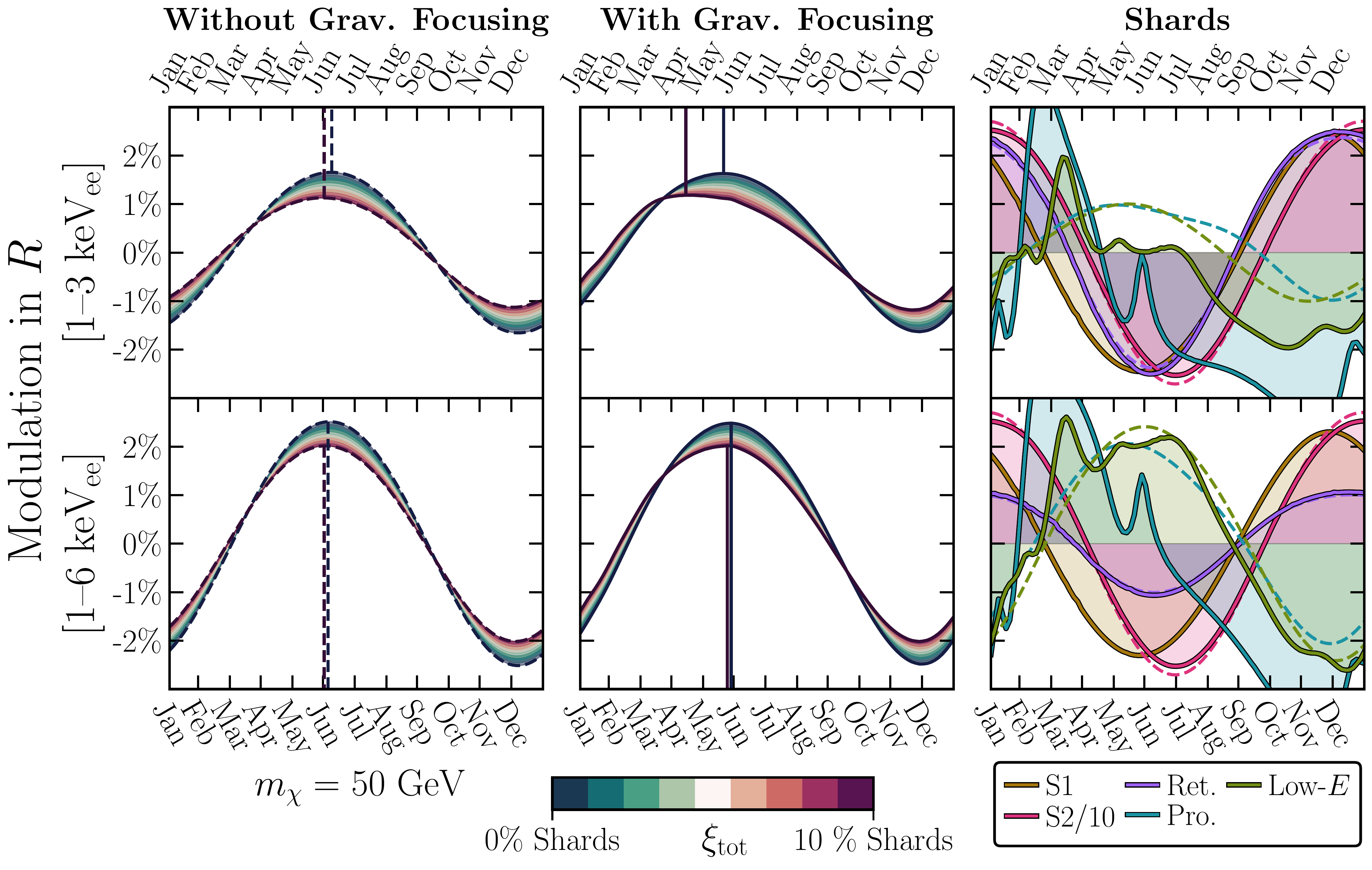}
  \caption{The modulation of the integrated event rate in two ionisation
    energy bins of 1--3 and 1--6 \keV (top and bottom panels in each
    column respectively). The WIMP particle model is kept the same from
    the previous Figure. The left-hand column shows the calculation
    ignoring gravitational focusing whereas the middle column shows the
    nonsinusoidal modifications that it gives rise to when it is
    included. The right-hand column shows each Dark Shard contribution
    individually. For visibility here we have suppressed the S2
    modulation by a factor of 10 as it is by far the dominant
    contribution for this DM mass and for these energy
    bins.}\label{fig:Shards_annualmod_binned}
  \end{figure*}

  The results presented so far have been time-independent. However,
  there are several classes of experiments for which the annual
  modulation brought about by the Earth's motion around the Sun is the
  primary signal~\cite{Drukier:1986tm,Freese:1987wu}. As the velocity
  distributions of the Dark Shards do not have an average velocity
  $\mathbf{v}=0$ in the galactic rest frame, they will distort
  modulation signals relative to the rest of the halo that does have
  zero average velocity.

  To obtain the modulation signal for a detector in a laboratory on
  Earth, $f_{\rm{lab}}(\mathbf{v},t)$, two effects have to be taken into
  account. The first is due to the time dependence in the Earth
  velocity, which can be written as,
  \begin{equation}
  \mathbf{v}_\oplus(t) = v_\oplus \bigg[ \cos{\left[\omega(t-t_a)\right]}\, \epsa + \sin{\left[\omega (t-t_a)\right]} \,\epsb \bigg].
  \end{equation}
  The Earth's velocity oscillates with frequency $\omega = 2\pi/(1\,{\rm
    year})$, $t_a\simeq 22$~March is the Vernal Equinox, $v_\oplus =
  29.79~\mathrm{km}~\!\mathrm{s}^{-1}$ is the average speed, and
  \begin{align}\label{eq:orbitvectors}
  \epsa &= (0.9941, 0.1088, 0.0042)  \, , \\ 
  \epsb &= (-0.0504, 0.4946, -0.8677) \, ,
  \end{align}
  are orthogonal unit vectors for the Earth's orbital plane expressed in galactic coordinates. For brevity, we have suppressed the corrections from the eccentricity
  of the Earth's orbit, but they are included up to second order in our
  numerical results.

  The second effect, known as gravitational
  focusing~\cite{Griest:1987vc,Sikivie:2002bj,Alenazi:2006wu,Lee:2013wza,Bozorgnia:2014dqa,DelNobile:2015nua},
  takes into account the Sun's gravitational influence on DM
  particles in the Solar System. For particles blowing through the Solar
  System, the Sun acts as a gravitational lens that enhances the flux in
  March relative to half a year later in
  September~\cite{Griest:1987vc,Lee:2013wza}. This effect is out of
  phase with the annual modulation induced by the Earth's rotation
  $\mathbf{v}_\oplus(t)$, which enhances the flux in June and reduces it
  in December~\cite{Freese:1987wu}. Ultimately, gravitational focusing is
  negligible for time-averaged
  signals~\cite{Alenazi:2006wu}, however it can be an important
  modification to consider for experiments that make use of timing
  information~\cite{Lee:2013wza}.

  When both effects are included, the time-dependent lab-frame velocity
  distribution is
  \begin{equation}
  f_{\rm{lab}}(\mathbf{v},t)=f_{\rm{gal}} (\mathbf{v}_\odot +  \mathbf{v}_\infty[\mathbf{v}_\oplus(t) + \mathbf{v}])\;.
  \end{equation}
  The function $\mathbf{v}_\infty[\mathbf{v}]$ gives the velocity that a
  particle would have had at infinity to have fallen into the Solar
  System with the velocity $\mathbf{v}$. Reference~\cite{Alenazi:2006wu}
  showed that it can be expressed as
  \begin{equation}
  \mathbf{v}_\infty[\mathbf{v}] = \frac{v_\infty^2 \mathbf{v}+\frac{1}{2}v_\infty u_{\rm esc}^2 \hat{\mathbf{x}}_\oplus(t) - v_\infty \mathbf{v}(\mathbf{v}\cdot \hat{\mathbf{x}}_\oplus(t))}{v_\infty^2 + \frac{1}{2} u_{\rm esc}^2 - v_\infty(\mathbf{v}\cdot \hat{\mathbf{x}}_\oplus(t))}\;,
  \end{equation}
  where $u_{\rm esc} = \sqrt{2GM_\odot/r_\oplus(t)}\simeq 42.3 \kms$,
  $v_\infty^2 = |\mathbf{v}|^2 - u_{\rm esc}^2$ and
  $\hat{\mathbf{x}}_\oplus(t) = -\sin{\left[\omega(t-t_a)\right]}\, \epsa
 + \cos{\left[\omega (t-t_a)\right]\, \epsb}$. In the limit that
  $M_\odot \to 0$, $\mathbf{v}_\infty[\mathbf{v}] \to \mathbf{v}$ and we
  recover the usual result in the absence of gravitational focusing.

  To highlight the impact of the Dark Shards on modulation signals, we
  focus on the best known experiments searching for it; namely DAMA/LIBRA~\cite{Bernabei:2013xsa,Bernabei:2018yyw}, together with the
  experiments trying to replicate DAMA/LIBRA's result by employing
  similar technology and NaI(Tl) crystals.  These include
  DM-Ice~\cite{deSouza:2016fxg}, KIMS~\cite{Kim:2018wcl},
  SABRE~\cite{Froborg:2016ova},
  Cosine~\cite{Adhikari:2018ljm,Adhikari:2019off} and
  ANAIS~\cite{Amare:2018sxx, Amare:2019jul}.  We do not provide a
  detailed fit to the DAMA/LIBRA data but instead focus our discussion
  on the general features introduced by the Dark Shards.  
  
  For more than a decade, the DAMA/LIBRA experiment has measured a persistent annually
  modulating signal above known backgrounds (however, see
  Refs.~\cite{McKinsey:2018xdb, Ferenc:2019esv}) with a significance in
  excess of~$9\sigma$~\cite{Bernabei:2018yyw}.  Although there is
  difficulty in explaining this signal with galactic DM while remaining
  consistent with other experiments, the signal does have many of the
  properties expected from galactic DM.  ANAIS~\cite{Amare:2018sxx, Amare:2019jul} and Cosine~\cite{Adhikari:2018ljm,Adhikari:2019off} have released
  first results and do not observe a statistically significant modulation, but they still require a few years of exposure to definitively
  test the DAMA/LIBRA result.

  To calculate the observable modulation signal for DAMA/LIBRA and the
  other NaI(Tl)-based experiments, we need to convert from the recoil
  energy $E_r$ to the ionisation energy $E_{\rm ee}$. They are related
  by $E_{\rm ee} = Q E_r$, where $Q$ is the quenching factor.
  DAMA/LIBRA have historically used the values $Q = 0.3$ and $0.09$ for
  Na and I respectively~\cite{Bernabei:1996vj}, but we use the lower
  value $Q=0.1$ for Na to reflect more recent
  measurements~\cite{Joo:2018hom}.  To account for the finite PMT
  resolution, we convolve the rate by a Gaussian energy-dependent energy
  resolution with width, $\sigma_{\rm{E}}(E_{\rm ee}) = \alpha \sqrt{E_{\rm ee}} +
  \beta E_{\rm ee}$, where $\alpha = 0.448 \, \sqrt{\rm \keV_{ee}}$ and
  $\beta = 9.1 \times 10^{-3}$.

  In Fig.~\ref{fig:Shards_annualmod_gravfocus}, we show one particular measure of
  modulation amplitude as a function of ionisation energy for a 50~\GeV
  particle.  The modulation amplitude is calculated by taking the
  difference of the differential recoil event rate on June 2 and Dec 2
  (the days when the differential rate is maximised/minimised in the
  Standard Halo Model).  This amplitude exhibits the well-known phase flip, here at
  $E_{\rm ee} \approx 1.3~\keV_{\rm ee}$.  As in previous figures,
  the effect of the Dark Shards is indicated by the shading in
  Fig.~\ref{fig:Shards_annualmod_gravfocus}.  We see that as
  $\xi_{\rm{tot}}$ increases (i.e.,\ moves from green to red), the
  amplitude of the modulation amplitude decreases in the window between~1
  and~$3~\keV_{\rm ee}$ and in the grey shaded region below $1~\keV_{\rm
    ee}$ (which is below the energy threshold set for DAMA/LIBRA phase-2).

  \begin{figure*}[t]
  \centering
  \includegraphics[width=\textwidth,trim={0cm 0cm 0cm 0cm},clip]{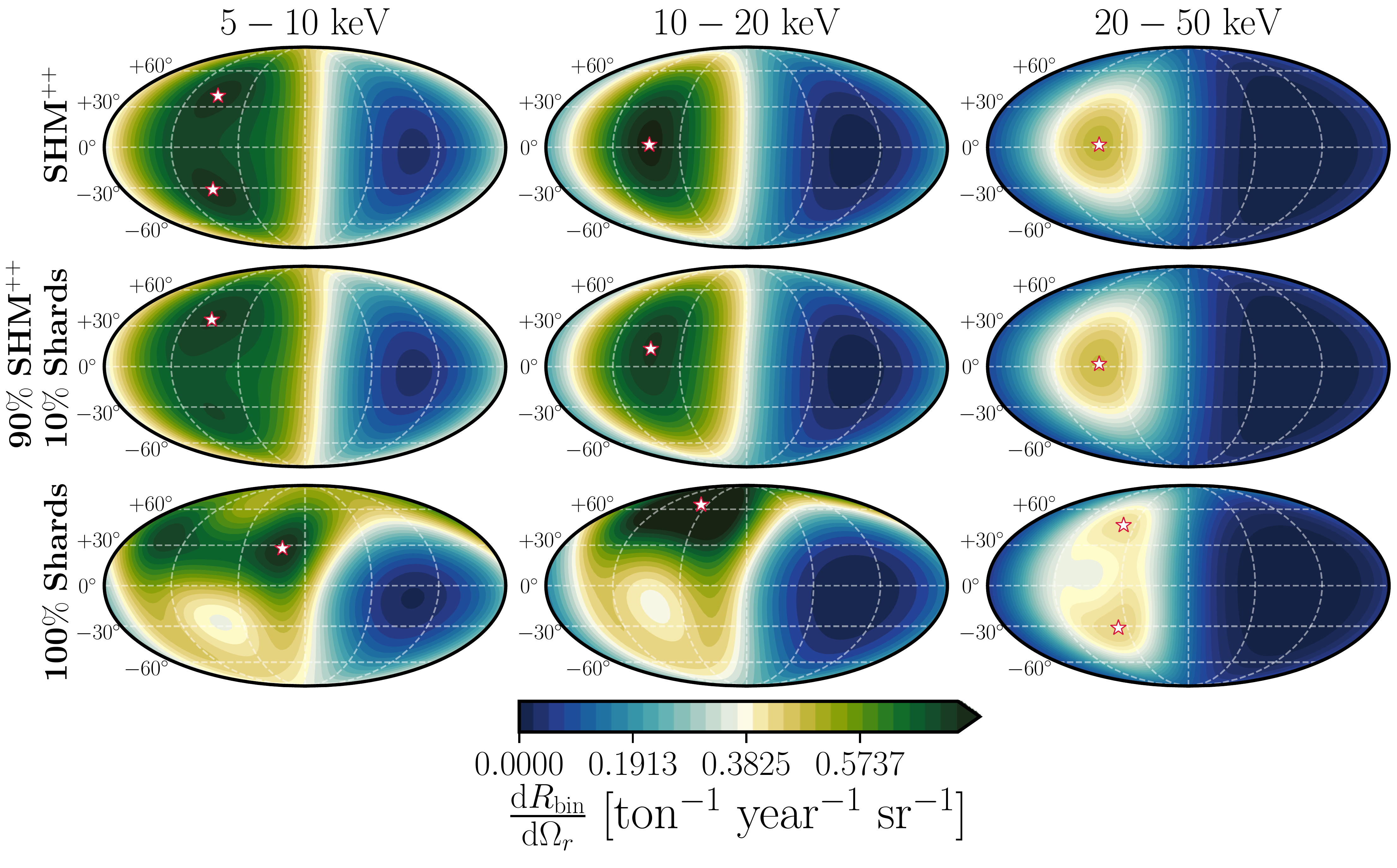}
  \caption{Mollweide projection in galactic coordinates of the value of
    the angular recoil rate, integrated over the recoil energy bins displayed at the top of the figure. We assume a 100 \GeV DM mass with $\sigma^{\rm SI}_p =
    10^{-46}$ cm$^2$ scattering with~$^{19}$F. The upper three panels
    show the angular distribution of the SHM$^{++}$, the middle panels
    show the distribution after the inclusion of a 10\% contribution
    from the Dark Shards and the lower panels show just the Shards
    distribution (equivalent to $\xi_{\rm tot} = 1$). In each case we account for an
    angular resolution of 20$^\circ$. The peak
    direction or directions are indicated with white stars.}\label{fig:directional}
  \end{figure*}

  The coloured lines in Fig.~\ref{fig:Shards_annualmod_gravfocus} show
  the modulation amplitude for the individual Dark Shard categories. The coloured lines
  make it clear that S2 leads to the most significant change.
   S2 has a large modulation amplitude and
  the phase flip is reversed relative to the other Dark Shards and the
  smooth components of the halo: for S2 the rate in June is smaller at
  low energies while the rate in December is larger at higher energies.
  The dominant direction of S2 is towards negative galactic-$Z$
  (cf.\ Fig.~\ref{fig:XYZ}) so the scattering rate at higher energies
  will be largest when the Earth's velocity is moving fastest in
  positive~$Z$. For the Earth's orbit, this occurs in
  December, explaining the behaviour in
  Fig.~\ref{fig:Shards_annualmod_gravfocus}.  None of the other Dark
  Shards move on orbits so distinct from the smooth halo component
  (cf.\ Fig.~\ref{fig:Shards_DMFlux}), which is why their modulation
  behaviour is similar to the smooth halo component.

  The dashed and solid lines in
  Fig.~\ref{fig:Shards_annualmod_gravfocus} show the modulation
  amplitude without and with gravitational focusing. The dashed and
  solid lines largely overlap showing that the effect of focusing on
  this measure of the modulation is small.  
  
  To make the impact of gravitational focusing more apparent, we show
  in Fig.~\ref{fig:Shards_annualmod_binned} the time-dependent
  component of the rate over the course of one year for the same DM
  mass and cross section used in
  Fig.~\ref{fig:Shards_annualmod_gravfocus}.  The rate versus time in
  Fig.~\ref{fig:Shards_annualmod_binned} has been integrated over two
  recoil energy bins of 1--3 and 1--6~$\keV_{\rm ee}$ (aligning with
  results often presented by DAMA/LIBRA).  The left-most panels show
  the rates without focusing, which are clean sinusoidals. The main
  effect of the Dark Shards is to reduce the modulation amplitude,
  whereas the day when the rate is maximum (indicated by the dashed
  vertical lines) is changed only slightly.  The middle panels show
  the rates when gravitational focusing is included. For the
  1--3~$\keV_{\rm ee}$ bin in particular, the effect is large: the
  peak day shifts by more than one month and the sinusoid is
  distorted.  The right panels show the effects from the individual
  Dark Shards.  The dominant component causing this change is S2 (in
  fact the modulation of S2 has been suppressed by a factor of ten in
  the right-hand column of Fig.~\ref{fig:Shards_annualmod_binned} so
  that all categories are visible).  The structure of S2 is disjoint
  in velocity space, with a prominent wrap at negative vertical
  velocity and a subsidiary wrap at positive
  velocity. This is also discernible in the middle panel
  of Fig.~\ref{fig:XYZ}. The gravitational focusing of the annual modulation signal is particularly
  affected by the orientation of the subsidiary wrap, (S2b in Table~\ref{tab:shards}) which roughly
  lies in the plane of the ecliptic. This is an intriguing result, however some doubt has been cast in the literature~\cite{Yu19} as to whether this additional wrap is truly a component of S2, or merely a collection of stars lying at a coinciding with S2 in action space. Ongoing work will address this issue in the near-future.

  \begin{figure*}[t]
  \centering
  \includegraphics[width=\textwidth,trim={0cm 0cm 0cm 0cm},clip]{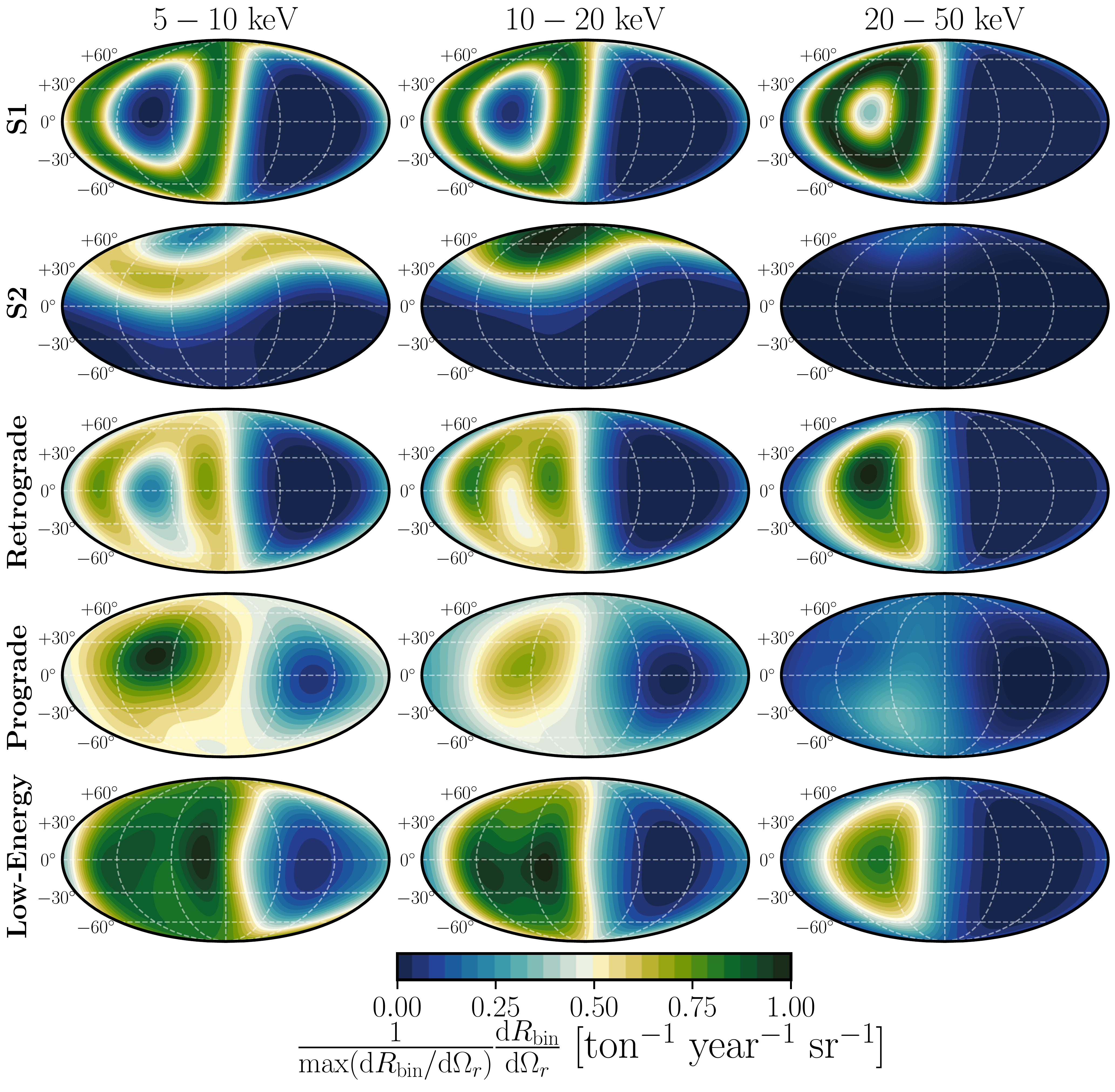}
  \caption{As in the bottom row of Fig.~\ref{fig:directional} but with each Shard category shown
    individually. Since S1 and S2 are streams focused around single directions, their recoil distributions appear as rings with recoil energy-dependent angular radii. The Prograde, Retrograde and Low Energy categories have multiple components so lead to more complex topologies. Lower energy recoils are associated with larger angles away from Cygnus; recoil directions from Low Energy and Prograde Shards in particular extend much more widely over the sky.}\label{fig:directional_indiv}
  \end{figure*}

  In light of this result, we reiterate the conclusion of
  Sec.~\ref{sec:axions} that the S2 stream is surprisingly important for
  some DM signals and deserves to be studied in much greater
  depth. Furthermore, the comparison of
  Figs.~\ref{fig:Shards_annualmod_binned}
  and~\ref{fig:Shards_annualmod_gravfocus} inspires us to urge that data
  from annual modulation searches be analysed in a framework that allows
  for higher harmonics (this requires of course that energy and timing
  information on recoil events are made available). Many of the
  interesting effects displayed in
  Fig.~\ref{fig:Shards_annualmod_binned} are missed when calculating a
  modulation amplitude which assumes a single sinusoid with a fixed peak
  day. Whereas they could be picked up by a Fourier mode
  decomposition. We have left an exercise in such a calculation in the
  GitHub repository attached to this
  paper~\cite{DarkShards}, but in
  the interest of keeping this section concise we do not discuss this
  further.

  \subsection{Directional signals}\label{sec:directional}

  We showed in Fig.~\ref{fig:Shards_DMFlux} that the angular dependence
  of the DM flux exhibits marked differences in our model compared with
  the expectation from a smooth halo. Many classes of experiment
  searching for DM are sensitive to this directional dependence of the
  underlying signal, as already discussed in the context of axions in
  Sec.~\ref{sec:axiondirection}. Now we turn to the directional
  dependence of nuclear recoil signals.

  The detection of the directionality of nuclear recoils is a promising
  technique for the discovery of WIMP-like
  particles~\cite{Mayet:2016zxu}. The galactic anisotropy of the
  DM flux cannot be mimicked by any known
  terrestrial~\cite{Leyton:2017tza} or cosmic
  background~\cite{Mei:2005gm}, including solar
  neutrinos~\cite{O'Hare:2015mda, OHare:2017rag}. 
  However, measuring the recoil
  direction for \keV-scale recoil tracks is experimentally
  challenging. In liquid or solid state detectors, recoil tracks are
  typically $\mathcal{O}(\mathrm{nm})$, whereas in gas they can be
  $\mathcal{O}(\mathrm{mm})$. This means that a directional detector
  requires either a readout method with high spatial resolution
  (such as x-ray imaging of nuclear emulsions~\cite{Aleksandrov:2016fyr,Agafonova:2017ajg}) or detection
  media with very low pressure. We focus on the latter technique here since low-pressure gas time projection chambers (TPCs) have seen many more years of consistent progress in their experimental realisation, see e.g.,~Refs.~\cite{Santos:2011kf,Riffard:2013psa,Nakamura:2015iza,Monroe:2011er,Leyton:2016nit,Daw:2011wq,Battat:2014van,Battat:2016xaw,CYGNO:2019aqp}).
Most recently the design for a tonne-scale network of directional
  gas-TPCs named {\sc Cygnus}~\cite{CYGNUS} has been put forward. The
  preliminary design study for {\sc Cygnus} is for a gaseous time projection
  chamber with a total active volume between 1000~m$^3$ and
  100,000~m$^3$, using a He:SF${_6}$ gas mixture at atmospheric pressure~\cite{CYGNUS}.

  Directional nuclear recoil detectors such as {\sc Cygnus} aim to measure the
  direction~$\hat{\textbf{q}}$ associated with each nuclear recoil in
  addition to the nuclear recoil energy~$E_r$. The double differential
  event rate for spin independent elastic scattering as a function of recoil
  energy, recoil direction and time~is,
  \begin{equation}\label{eq:finaleventrate-directional}
   \frac{\textrm{d}^2R(t)}{\textrm{d}E_r\textrm{d}\Omega_r} =\frac{1}{2\pi}\frac{\rho_0}{m_\chi} \frac{A^2 \sigma^{\mathrm{SI}}_{p} }{2 \mu_{p}^2} F^2(E_r) \, \hat{f}(\vmin,\hat{\textbf{q}},t) \, .
  \end{equation}
  This formula is similar to the non-directional rate for
  spin independent scattering, except for the factors of $1/2\pi$ and
  $\hat{f}(\vmin,\hat{\textbf{q}},t)$ instead of $g(\vmin,t)$. This is
  the ``Radon transform'' of the velocity
  distribution~\cite{Gondolo:2002np,Radon}:
  \begin{equation}
   \hat{f}(\vmin,\hat{\textbf{q}},t) = \int \delta\left(\textbf{v} \cdot \hat{\textbf{q}} - \vmin\right) f_{\rm{lab}}(\textbf{v},t)\, \textrm{d}^3 v\, .
  \end{equation}
  We show the angular recoil distributions integrated over three recoil
  energies bins in Fig.~\ref{fig:directional}. We assume
  spin independent scattering with $m_{\chi}=100$~\GeV and a fluorine
  target, and apply a Gaussian angular resolution of 20$^\circ$.  We compare the
  recoil maps from the SHM$^{++}$ which contains only the smooth
  components of the halo (top row), the halo model with Dark Shards
  where $\xi_{\rm tot} = 10\%$ (middle row), and the recoil maps from
  the Dark Shards alone (bottom row).
  Figure~\ref{fig:directional_indiv} can be used to observe each of the categories
  individually for comparison.

  Over all energy bins in Fig.~\ref{fig:directional}, the rate peaks in
  the forward half of the sky. One interesting and prominent feature in
  the top row (where $\xi_{\rm tot}=0$) is that the rate has two maxima at
  positive and negative galactic latitudes of $\pm 60^\circ$. This is due to the influence of the radially anisotropic Sausage. 
  The intuition behind this feature is subtle, but can be thought of as a 
  manifestation of the ring feature originally discussed in the context of 
  directional detectors in Ref.~\cite{Bozorgnia:2011vc}. The ring feature 
  shows up at low recoil energies for large DM masses and isotropic 
  halo models. In a certain range of low energies, the Radon transform 
  is not maximised at $\hat{\textbf{q}} = -\hat{\textbf{v}}_\textrm{lab}$ 
  as it is at high energies, but rather at an energy-dependent angle away 
  from~$\vlab$. This leads to recoils being focused along a ring around~$\vlab$. 
  For the radially anisotropic Sausage there are instead two rings centred 
  at wide angles away from~$\vlab$. The centres of the rings are closer 
  to the galactic radial axis but do not precisely align with it because of 
  the boost into the lab frame. When we calculate the recoil map for 
  the Sausage across the sky, the rate is maximised when the two rings 
  intersect. Because the ring centres are equal angular distances away 
  from~$\vlab$ in galactic longitude, the intersections occur at galactic 
  latitudes above and below $\vlab$, hence the position of the two stars 
  in the top panel of Fig.~\ref{fig:directional}.
  
  The influence of the Shards when included as a part of this distribution 
  is to distort the symmetry of these two maxima about the galactic plane. 
  The Shards also present several interesting effects when comparing the 
  distribution across the sky between bins in recoil energy. Each Shard presents 
  its own ring feature centred around the direction of its velocity. Kinematically 
  a nuclear recoil will scatter at smaller angles away from the direction 
  of the impinging particle, and in this context we see that as the angular 
  radii of each ring becoming smaller as recoil energy increases. This can 
  be seen in the panels for the S1 and S2 streams in Fig.~\ref{fig:directional_indiv}.
  
  The impact of the Shards is notable departures from the central assumption 
  in directional detection, which is that the flux is rotationally symmetric around the Cygnus constellation.
 The overall forward-backward anisotropy is however largely preserve, despite
  the fact that some of the Low Energy and Prograde categories have sizeable
  event rates in the backward half of the sky (see Fig.~\ref{fig:directional_indiv}).  
  If the anisotropy were reduced, this could
   harm the powerful discovery capabilities of directional WIMP 
   searches~\cite{Mayet:2016zxu}, so it is comforting that this is not the case.
 The anisotropy in fact seems to be quite strongly peaked in the forward direction 
 when we go to the higher energy bins. This is largely because the Shards that 
 have the largest angles away from Cygnus do not have recoils scattering at 
 higher energies. It is only the retrograde substructures that align with 
 Cygnus that are still present at high recoil energies and therefore add to the anisotropy.
 
The sum of all the Dark Shards would likely require many hundreds of 
events to distinguish from a smooth DM halo distribution. Nevertheless, in 
the context of WIMP-like dark matter, the measurement of the directionality 
of nuclear recoils would likely be the only feasible way to probe this part of 
the DM halo directly. Fewer events would be required in the cases of single 
streams such as S1 and S2, which lead to much more striking changes to the 
expected directional signals. To quantify these statements, we can 
estimate the number of events required in a benchmark directional detector
 to measure a~$3\sigma$ dipole anisotropy in the rates, $\mathcal{A}_{12} = R_1/R_2$, 
 between two hemispheres, labelled ``1'' and ``2'':
\begin{equation}
N_{\rm aniso} \approx \left( 3 \, \frac{R_1 + R_2}{R_1 - R_2} \right)^2 \, .
\end{equation}
The formula arises from simply assuming that the contrast in the 
rates, proportional to $(R_1-R_2)$, between the two hemispheres should be 
greater than a~3$\sigma$ deviation in the total number, proportional to $ 3\sqrt{R_1+R_2}$,
in order to measure the anisotropy.

We assume two orthogonal pairs of hemispheres, a ``forward-backward'' pair, 
which align with~$\vlab$, and an ``up-down'' pair, which is aligned $90^\circ$ 
from the forward-backward hemispheres, and approximately align with the Galactic 
North Pole. Taking a typical directional detector model assuming $^{19}$F 
recoils above a threshold of 3~keV, under the SHM$^{++}$ the forward-backward 
anisotropy is on the order of $\mathcal{A}_{\rm fw}\sim 5$ (requiring around 20 events to detect) 
and there is no up-down anisotropy, so $\mathcal{A}_{\rm ud}=1$. We find that over all 
detectable values of the DM mass, the Dark Shards in total give a negligible change to 
both $\mathcal{A}_{\rm fw}$ and $\mathcal{A}_{\rm ud}$. So the forward-backward 
dipole is largely preserved, but the slight up-down asymmetry caused by the Dark Shards 
would need in excess of $N_{\rm aniso}\simeq 1000$~events for $\xi_{\rm tot} \lesssim 50\%$.

Isolating S1 and S2, we find that they give rise to opposing effects, which reflect 
their perpendicular velocities. S1 enhances the forward-backward anisotropy, 
approximately linearly with $\xi_{\rm S1}$, whereas S2 enhances the 
up-down anisotropy with $\xi_{\rm S2}$. For 10\% of the local density in S1, we find 
that $\mathcal{A}_{\rm fw}$ is enhanced such that it would reduce the required 
number of events to detect it from 20 to 17; $\mathcal{A}_{\rm ud}$ is essentially 
unchanged. For 10\% in S2, the required number of events to detect $\mathcal{A}_{\rm fw}$ 
is only mildly increased from 20 to 23, but the up-down anisotropy is increased 
dramatically and is detectable with $N_{\rm aniso}\simeq 250$ events as opposed to $1000$. 
While these numbers are large, it is worth pointing out that in some ways they 
reflect upper limits. This simple procedure of calculating dipole anisotropies 
along two directions represents a model-independent test that could be
performed on directional data but it does not optimise for the particular direction 
of a given stream, or include the correlation with recoil energy information. A full likelihood-based 
analysis that includes this additional information would therefore require fewer events than we have estimated here.

\section{Summary}\label{sec:summary}

Thanks to \Gaia, the six-dimensional view of our local stellar halo is more extensive than ever before. The distributions of stars in action space are revealing much about the recent accretion history of the Milky Way. We have studied the properties of the substructure found in this distribution due to accretion during the {\it ex situ} phase of the Milky Way's formation. Substructure passing close by the Earth has implications for a wide range of experiments looking for DM.

The sample of stars we study here is derived from a cross-matching of the SDSS and \Gaia catalogues. Taking inspiration from the kinematic structure of the stellar halo in this sample, we began by describing the two smooth components present in the dark halo: a roundish, isotropic part and a radially anisotropic ``sausage''-shaped part. To this, we then introduced substructures. We considered in particular the large collection of action space substructures, or ``Shards'', present in the \Gaia-SDSS stellar halo sample. We individually modelled the Shards, identified those passing closest to the solar neighbourhood, and then used the models to build an approximate description for the collection of their DM counterparts: the ``Dark Shards''. We organised the Dark Shards into five categories in terms of overall significance and properties: S1, S2, Retrograde, Prograde, and Low Energy. When accounting for the Dark Shards, the resulting DM velocity and speed distributions contain notable non-Gaussian departures from their conventionally assumed Gaussian and Maxwellian forms (see Figs.~\ref{fig:Shards_fvgal} and~\ref{fig:Shards_fv_lab}).

Of the long list of Shards, we highlight the S1 and S2 streams as the most important ones nearby. They are present at highest significance in action space~\cite{Myeong:Preprint,Myeong:2017} and have the most important experimental implications. S1 has a very high Earth-frame speed of $\sim 550 \kms$, so is important for signals which are sensitive to the tail of the speed distribution. Several other substructures on retrograde orbits lead to phenomenologically similar effects.  These may be observable in nuclear recoil searches for WIMPs, if their representation in the local DM density is relatively large~\cite{OHare:2018trr}.

S2 gives rise to a more complex array of new effects. Unlike~S1, it has not been studied until now; despite the fact that it also passes through the solar neighbourhood and is of roughly equal prominence as an action space substructure in \Gaia-SDSS. S2 moves on a prograde, but highly polar orbit, so its DM enters the Milky Way disk from above. In the Earth frame, it contributes near the peak of the speed distribution at around $300 \kms$ (see Fig.~\ref{fig:axionspectrum}). Being present at lower speeds than S1 means that it can potentially impact signals in axion haloscopes in a much more remarkable way. In fact there are important implications for the very detection of the axion. Since sharper peaks show up more strongly over thermal noise, S2 is positioned at the ideal frequency to enhance the sensitivity of a haloscope to the axion-photon coupling. 

Streams with polar orbits imply peculiar modifications to the standard lore for time dependent signals induced by DM. Not only does S2 (in combination with several other Shards) shift the peak day of the annual modulation of the event rate of nuclear recoils, it also causes that modulation to become nonsinusoidal. Here, we have also included the often-neglected (but present) effect of gravitational focusing. Since this largely involves the modulation of the low-speed tail of the speed distribution, it is more important to take into account when considering the S2, Prograde and Low Energy Shards. We studied annual modulation in the context of NaI(Tl) experiments like DAMA/LIBRA, but these broad statements apply to the annual modulation of WIMP-like DM in general.

All of the Dark Shards have interesting and potentially detectable effects on angular and directional signals. Owing to the large quantity of retrograde material, the DM flux is enhanced in the direction of galactic rotation (towards the Cygnus constellation). This is to the benefit of direction-sensitive experiments.  However, the Dark Shards on radial or polar orbits lead to additional peaks in the DM flux that have large angular separation from the expected peak towards Cygnus (compare the two panels of Fig.~\ref{fig:Shards_DMFlux}). For nuclear recoil-based experiments like {\sc Cygnus} that aim to measure this directionality, the process of elastic scattering and the finite angular resolution partially washes out this substructure. 
But for direction-sensitive axion searches like CASPEr and QUAX, the modifications are  complex enough to warrant a dedicated study.

This paper is the first to sketch out some of the implications of the abundant substructure found by \Gaia. This field of activity is likely to become more important with future data releases, as the characterisation of stellar substructures will improve over the coming years. Perhaps the most important area for future study is the relationship between stellar substructures and their associated DM components. Despite some early progress and speculation in the literature, this problem remains underexplored and is ripe for further investigation.

\acknowledgments
C.A.J.O. is supported by the Grant No. FPA2015-65745-P from the Spanish Ministry of Economy and Business and the European Regional Development Fund. 
C.M. is supported by the Science and Technology Facilities Council (STFC) Grant No. ST/N004663/1. 
G.C.M. thanks the Boustany Foundation, Cambridge Commonwealth, European \& International Trust and Isaac Newton Studentship for their support of his work. 
The research leading to these results has received funding from the European Research Council under the European Union's Seventh Framework Programme (FP/2007-2013) / ERC Grant Agreement No.\ 308024.

\appendix
\section{Low Energy Shards}\label{sec:lowEshards}
\begin{table*}[t!] 
\ra{1.3}
\begin{tabularx}{0.95\textwidth}{c|cccccccc}
\hline\hline
{\bf Name}      & Number & $(X,Y,Z)$ & $(\Delta X,\Delta Y,\Delta Z)$ & $(v_R,v_\phi,v_z)$ & $(\sigma_R,\sigma_\phi,\sigma_z) $ & \quad $\langle [{\rm Fe}/{\rm H}] \rangle$ \quad        & \quad$P(\mathbf{x}_\odot)$\\
      &  of stars & kpc & kpc \quad\quad& $\kms$   \quad    & $\kms $ \quad &    &   $(\sigma)$  & \\
    \hline
NCand0   & 19        & $(9.4,-0.6,2.4)$  & $(1.6,2.1,3.0)$                      & $(364.6,94.2,56.3)$      & $(27.2,15.7,41.1)$                & $-1.4\pm 0.2$ & 0.5 \\
NCand2a  & 13        & $(9.6,-0.4,3.0)$  & $(1.1,2.3,3.5)$                      & $(270.2,133.8,4.8)$      & $(59.5,14.3,163.4)$               & $-1.6\pm 0.2$ & 0.9 \\
NCand2b  & 22        & $(10.0,0.3,2.3)$  & $(1.8,1.7,4.3)$                      & $(-250.8,139.5,-28.1)$   & $(69.6,14.9,160.7)$               & $-1.6\pm 0.2$ & 0.6 \\
NCand3   & 9         & $(9.4,0.5,-0.3)$  & $(1.3,2.1,2.7)$                      & $(-212.9,104.5,123.8)$   & $(30.8,8.0,192.7)$                & $-1.4\pm 0.1$ & 0.8 \\
NCand6a  & 11        & $(8.1,0.2,1.1)$   & $(0.9,1.1,1.5)$                      & $(-11.9,-32.9,177.6)$    & $(30.0,13.5,7.2)$                 & $-2.0\pm 0.1$ & 0.2 \\
NCand6b  & 10        & $(7.5,1.4,1.7)$   & $(1.0,1.6,1.4)$                      & $(-3.6,-30.4,-180.5)$    & $(30.2,11.4,10.8)$                & $-2.0\pm 0.3$ & 0.6 \\
NCand7   & 11        & $(8.4,1.2,0.7)$   & $(2.7,1.7,4.6)$                      & $(-178.5,-114.8,-107.3)$ & $(188.3,14.6,302.6)$              & $-1.3\pm 0.3$ & 0.8 \\
NCand8   & 7         & $(10.1,-1.3,2.6)$ & $(1.3,1.9,4.4)$                      & $(26.2,375.0,-34.1)$     & $(47.9,12.8,85.4)$                & $-1.8\pm 0.2$ & 1.0 \\
NCand9a  & 9         & $(6.9,0.6,2.4)$   & $(1.3,0.9,2.7)$                      & $(-0.8,64.9,-189.3)$     & $(24.1,13.9,20.9)$                & $-1.8\pm 0.4$ & 0.9 \\
NCand9b  & 13        & $(7.9,0.2,2.3)$   & $(1.7,1.2,3.4)$                      & $(-12.2,66.8,190.6)$     & $(29.7,16.4,28.1)$                & $-2.1\pm 0.3$ & 0.2 \\
NCand10a & 23        & $(9.0,-0.4,1.5)$  & $(1.6,1.8,3.3)$                      & $(-21.9,34.2,233.1)$     & $(61.7,16.8,43.9)$                & $-1.9\pm 0.3$ & 0.1 \\
NCand10b & 13        & $(9.1,0.0,0.9)$   & $(1.7,1.6,4.2)$                      & $(-1.5,23.9,-186.2)$     & $(60.8,19.8,41.7)$                & $-1.8\pm 0.4$ & 0.2 \\
NCand11  & 20        & $(9.6,0.9,1.5)$   & $(2.4,2.4,4.9)$                      & $(-8.5,-212.9,98.8)$     & $(72.3,21.1,89.5)$                & $-1.8\pm 0.1$ & 0.4 \\
NCand12  & 18        & $(10.3,0.7,1.7)$  & $(2.0,1.8,5.1)$                      & $(1.9,128.2,-152.3)$     & $(137.9,12.7,126.9)$              & $-1.7\pm 0.3$ & 0.4 \\
NCand13a & 12        & $(7.1,0.2,1.6)$   & $(1.3,1.3,1.6)$                      & $(7.9,144.6,141.2)$      & $(35.7,17.2,24.6)$                & $-1.7\pm 0.4$ & 0.7 \\
NCand13b & 17        & $(7.4,0.7,1.0)$   & $(0.7,0.7,1.7)$                      & $(-0.1,163.6,-137.5)$    & $(25.4,12.7,15.4)$                & $-1.7\pm 0.3$ & 1.1 \\
NCand16  & 5         & $(11.1,-0.3,1.9)$ & $(3.0,1.3,4.0)$                      & $(-75.1,-5.0,62.7)$      & $(92.3,6.6,227.0)$                & $-1.9\pm 0.5$ & 0.5 \\
NCand17  & 15        & $(10.5,0.7,2.7)$  & $(1.8,1.5,3.1)$                      & $(1.5,224.8,-18.9)$      & $(86.3,29.0,95.5)$                & $-1.7\pm 0.2$ & 1.4 \\
NCand18  & 11        & $(10.3,0.9,0.3)$  & $(1.6,1.5,2.8)$                      & $(71.0,-84.6,-338.4)$    & $(45.7,16.4,29.6)$                & $-2.2\pm 0.2$ & 0.6 \\
NCand19a & 13        & $(9.2,-0.2,0.9)$  & $(1.6,1.3,2.3)$                      & $(-11.9,-147.8,176.3)$   & $(51.3,17.1,24.1)$                & $-2.0\pm 0.2$ & 0.6 \\
NCand19b & 10        & $(7.6,0.7,3.9)$   & $(1.0,3.0,3.7)$                      & $(-1.2,-135.2,-99.6)$    & $(39.2,9.6,45.4)$                 & $-1.8\pm 0.2$ & 1.1 \\
NCand20  & 9         & $(8.0,0.1,1.8)$   & $(0.9,0.8,3.3)$                      & $(-34.7,-145.3,-92.9)$   & $(27.2,14.2,39.5)$                & $-2.0\pm 0.2$ & 0.1 \\
NCand21a & 15        & $(9.0,0.2,2.9)$   & $(2.2,2.2,3.5)$                      & $(-6.5,-117.2,226.9)$    & $(65.2,12.6,34.0)$                & $-1.7\pm 0.3$ & 0.3 \\
NCand21b & 11        & $(9.2,0.5,2.1)$   & $(1.6,2.6,1.9)$                      & $(-26.0,-113.5,-243.7)$  & $(59.0,10.6,23.3)$                & $-1.9\pm 0.2$ & 0.4 \\
NCand22  & 20        & $(9.7,-0.0,3.3)$  & $(2.4,1.8,4.0)$                      & $(-19.3,-174.6,-44.4)$   & $(95.5,17.0,127.2)$               & $-1.9\pm 0.2$ & 0.3 \\
NCand23  & 8         & $(7.2,-0.1,4.8)$  & $(2.1,1.7,2.6)$                      & $(-450.2,-25.4,78.9)$    & $(22.4,12.9,27.9)$                & $-1.4\pm 0.1$ & 1.0 \\
NCand24a & 7         & $(10.0,0.0,1.5)$  & $(2.3,2.0,2.5)$                      & $(-80.8,253.9,-78.9)$    & $(30.5,14.9,29.9)$                & $-1.8\pm 0.2$ & 0.5 \\
NCand24b & 8         & $(10.6,0.5,2.1)$  & $(2.6,0.6,2.3)$                      & $(120.3,269.2,1.6)$      & $(23.2,18.2,43.3)$                & $-1.8\pm 0.3$ & 0.8 \\
NCand25  & 8         & $(11.2,1.7,0.9)$  & $(1.7,2.8,3.4)$                      & $(-95.0,199.4,-20.1)$    & $(87.1,17.8,84.7)$                & $-1.4\pm 0.1$ & 1.8 \\
NCand26  & 7         & $(10.3,1.0,4.0)$  & $(2.2,1.7,5.4)$                      & $(9.2,212.6,183.7)$      & $(78.1,11.7,29.4)$                & $-1.5\pm 0.3$ & 0.5 \\
NCand27  & 17        & $(7.5,0.5,2.4)$   & $(1.4,1.6,2.3)$                      & $(-18.0,-108.3,-54.5)$   & $(40.7,22.2,123.9)$               & $-2.0\pm 0.2$ & 0.3 \\
NCand28a & 12        & $(9.4,1.1,1.6)$   & $(1.8,2.2,4.0)$                      & $(-299.4,28.1,-10.2)$    & $(25.9,5.2,61.5)$                 & $-1.4\pm 0.3$ & 0.8 \\
NCand28b & 8         & $(9.0,-0.3,3.4)$  & $(1.0,1.5,3.1)$                      & $(249.5,19.5,115.5)$     & $(67.6,7.6,118.8)$                & $-1.3\pm 0.1$ & 1.7 \\
NCand29  & 14        & $(9.2,0.9,1.2)$   & $(1.7,1.1,4.4)$                      & $(-73.0,-325.6,-113.6)$  & $(166.5,27.8,92.3)$               & $-1.8\pm 0.5$ & 0.7\\

\hline\hline
 \end{tabularx}
\caption{As Table~\ref{tab:shards} but for the Low Energy Shards category.} 
\label{tab:restofshards}
\end{table*}

In Table~\ref{tab:restofshards} we list the remaining Low Energy Shards that are included in our main results but not in Table~\ref{tab:shards}.

\maketitle
\flushbottom

\bibliographystyle{apsrev4-1}
\bibliography{DarkShards}

\end{document}